\documentclass[a4paper,11pt]{article}
\usepackage[printwatermark]{xwatermark}

\usepackage{graphicx}
\usepackage{lipsum}
\usepackage{tikz}
\usepackage{paralist}
\usepackage{amsmath}
\usepackage{amsfonts}
\usepackage{amssymb}
\usepackage{graphicx, rotating}
\usepackage{epstopdf}
\usepackage{epsfig}
\usepackage{latexsym}
\usepackage{multirow}
\usepackage{color}
\usepackage[top=2.8 cm, bottom=2.8 cm, left=3 cm, right=3 cm]{geometry}
\usepackage{amstext} 
\usepackage{array} 
\usepackage{tabularx}
\usepackage{authblk}
\usepackage{colortbl}
\usepackage{cite}
\usepackage{textcomp}
\definecolor{DarkGreen}{rgb}{0.0, 0.2, 0.13}

\pdfoutput=1
\usepackage{listings}
\newcolumntype{L}{>{$}l<{$}} 
\newcommand{\ba}{\begin{eqnarray}}
\newcommand{\ea}{\end{eqnarray}}
\newcommand{\like}{\mathcal{L}}

\newcommand{\DS}{DarkSide-50}
\newcommand{\be}{\begin{equation}}
\newcommand{\ee}{\end{equation}}

\newcommand*{\email}[1]{\href{mailto:#1}{\nolinkurl{#1}} } 
\usepackage{hyperref}
\hypersetup{
colorlinks = true,
linkcolor = black,
citecolor={blue}
}

\begin{document}

\title{\bf Migdal effect and photon Bremsstrahlung: improving the sensitivity to light dark matter of liquid argon experiments}

\author[1]{G.~Grilli~di~Cortona\footnote{\email{ggrillidc@fuw.edu.pl}}}
\author[2,3]{A.~Messina\footnote{\email{andrea.messina@uniroma1.it}}}
\author[2,3]{S.~Piacentini\footnote{\email{stefano.piacentini@uniroma1.it}}}

\affil[1]{Institute of Theoretical Physics\\ Faculty of Physics, University of Warsaw\\ ul. Pasteura 5, PL–02–093 Warsaw, Poland}
\affil[2]{Dipartimento di Fisica\\  
 Sapienza Universit\`a di Roma, I-00185, Italy} 
\affil[3]{Istituto Nazionale di Fisica Nucleare\\  
 Sezione di Roma, I-00185, Italy}

\date{\today}

\maketitle
\begin{abstract}
\noindent 
The search for dark matter weakly interacting massive particles with noble liquids has probed  masses down and below a GeV/$c^2$. The  ultimate limit is represented by the experimental threshold on the energy transfer to the nuclear recoil. Currently, the experimental sensitivity has reached a threshold equivalent to a few ionization electrons. 
In these conditions, the contribution of a Bremsstrahlung photon or a so-called Migdal electron due to the sudden acceleration of a nucleus after a collision might be sizable.
In the present work, we use a Bayesian approach to study how these effects can be exploited in experiments based on liquid argon detectors. In particular, taking inspiration from the \DS\ public spectra, we develop a simulated experiment to show how the Migdal electron and the  Bremsstrahlung photon allow to push the experimental sensitivity 
down to masses of 0.1~GeV$/c^2$, extending the search region for dark matter particles of previous results. For these masses we estimate the effect of the Earth shielding that, for strongly interacting dark matter, makes any detector blind. Finally, we show how the sensitivity scales for higher exposure.

\end{abstract}

\clearpage
\begingroup
\color{black} 
\tableofcontents
\endgroup

%%%%%%%%%%%%%%%%%%%%%%%%%%%%%%%%%%%%%%
\section{Introduction} 
%%%%%%%%%%%%%%%%%%%%%%%%%%%%%%%%%%%%%%

Although astronomical and cosmological observations strongly support the existence of dark matter (DM) \cite{1932BAN.....6..249O,1937ApJ....86..217Z,Rubin:1980zd,Aghanim:2018eyx}, its nature -- its mass and interactions with the Standard Model (SM) -- has not yet been revealed. In the past decades, a huge experimental effort has been developed to detect DM. This program has focused primarily on DM masses ${m_{\chi}\gtrsim 1{\rm\:GeV}/c^2}$, motivated by the explanation of the current abundance as a thermal relic in the form of weakly-interacting massive particles (WIMPs). Direct detection experiments \cite{Akerib:2016vxi,Aprile:2018dbl,Agnese:2018col,Ren:2018gyx,Agnes:2018fwg, Aprile:2019xxb, Amole:2019fdf,Ajaj:2019imk}, searching for dark matter induced nuclear recoils in underground detectors \cite{Goodman:1984dc}, are among the numerous experiments that have been built to detect these interactions. 
The DAMA/NaI and DAMA/LIBRA experiments have results \cite{Bernabei:2003za, Bernabei:2005hj, Bernabei:2008yi, Bernabei:2013xsa, Bernabei:2018yyw}  that are interpreted by the DAMA collaboration as a strong evidence for the presence of DM particles in the galactic halo. However, the failure of observing incontrovertible evidences of a DM signal may be interpreted as in tension with the WIMP paradigm. These results stimulated the effort of examining alternative signals to nuclear recoil, including dark matter scattering on electrons \cite{Essig:2011nj,Essig:2012yx,Hochberg:2015pha,Lee:2015qva,Essig:2015cda,Roberts:2016xfw,Emken:2017erx,Essig:2017kqs,Cavoto:2017otc,Bertuzzo:2017lwt,Agnes:2018ves,Abramoff:2019dfb,Catena:2019gfa, Hryczuk:2020trm} or secondary signals \cite{Kouvaris:2016afs,Ibe:2017yqa,Dolan:2017xbu,Bell:2019egg}, and motivated the interest in sub-GeV/$c^2$ DM \cite{Agnes:2018ves,Agnese:2018gze,Aguilar-Arevalo:2016ndq,Akerib:2018hck, Abdelhameed:2019hmk}. 

Direct detection experiments generically are insensitive to nuclear recoils with energy below the keV, corresponding to sub-GeV/$c^2$ dark matter scattering.  This relies on the assumption that the electron cloud around the nucleus follows instantaneously the nucleus itself, keeping the atom neutral. However, the sudden acceleration of a nucleus after a collision may lead to excitation and ionization of atomic electrons. This is an old idea from neutron-nucleus scattering experiments \cite{Migdal:1941,Landau:1990qp,RUIJGROK1983537,Vegh_1983,Baur_1983,Sharma:2017fmo}. Furthermore, the electron will get accelerated in order to follow the nuclear recoil trajectory, resulting in a finite probability that a photon will be emitted via Bremsstrahlung. Therefore, this new process may lead to energetic photons and ionization electrons produced from the primary interaction. The first process is the Bremsstrahlung photon emission from a nucleus \cite{Kouvaris:2016afs}, the latter is the Migdal effect \cite{Ibe:2017yqa,Dolan:2017xbu,Bell:2019egg}, and they both have been already exploited by experimental collaborations \cite{Akerib:2018hck,Armengaud:2019kfj,Liu:2019kzq,Aprile:2019jmx}. 

In the present work we examine both the Migdal effect and photon Bremsstrahlung from the nucleus in experiments exploiting LAr detectors, and we estimate the effect of the Earth atmosphere and crust that makes experiments blind to large cross sections. Previous DM searches
of this kind
include \DS\ \cite{Agnes:2018fwg} and DEAP-3600 \cite{Ajaj:2019imk}, and have mainly focused on masses greater than $10$~GeV/$c^2$. \DS\ has published a low-mass analysis \cite{Agnes:2018ves}, exploiting the ionization-only signal, sensitive down to masses of $1.8$~GeV/$c^2$. In this article, we show how this analysis could be extended down to masses of $0.1$~GeV/$c^2$ including signals from the Migdal effect and the photon Bremsstrahlung. 

The rest of the paper proceeds as follows. In section~\ref{sec:ME_PB} we review the Migdal effect and photon Bremsstrahlung process, show their differential rates in the LAr detector, and describe the effect of the Earth attenuation. Section \ref{sec:inference} reviews the analysis with LAr. We describe the probabilistic inference and all the details of our simplified treatment of systematic effects. In section \ref{sec:results}, we show the expected sensitivity and the projections for higher exposure.
Finally we conclude in section \ref{sec:conclusions}. In addition, we provide the numerical codes used to evaluate the Migdal and Bremsstrahlung rates \cite{ref:github_DDrates}, and to perform the statistical analysis \cite{ref:github_LAr-MigdalLimits}.

\section{Migdal effect and photon Bremsstrahlung}
\label{sec:ME_PB}

We start this section describing our notation and our assumptions for the elastic DM-nucleus scattering rates, and continue presenting the computation of the differential rates for the Migdal effect and the photon Bremsstrahlung process.

The elastic DM-nucleus differential rate with respect to the nuclear recoil energy $E_R$, per unit detector mass, is 
\be
\frac{dR_{NR}}{dE_R}=N_T\frac{\rho_\chi}{m_\chi}\int_{v>v_{\min}}\frac{d\sigma_{SI}}{dE_R}v f(\vec{v})d^3v,
\label{eq:dRNRdER}
\ee
here $N_T$ is the number of target nuclei per unit detector mass, $E_R$ is the recoil energy given by an incoming dark matter particle with velocity $v>v_{min}= \sqrt{(m_N E_R)/(2 \mu_N^2)}$, $m_\chi$ and $m_N$ are the DM and nucleus mass, respectively, while $\mu_{i}=m_i m_\chi/(m_i+m_\chi)$ is the reduced mass of the nucleus or nucleon-DM system (with $i=N,p$). The rate depends on our assumptions on the local dark matter density, $\rho_\chi$, and the dark matter velocity distribution $f(\vec{v})$. 
In this work, we use the value\footnote{New determinations of the local dark matter density give results that fall in the range $\sim(0.3-0.4)\,\,\mathrm{GeV}/c^2/\mathrm{cm}^3$, see \cite{Eilers_2019,Karukes:2019jxv,deSalas:2019pee,Cautun:2019eaf}.} $\rho_\chi=0.3\,\,\mathrm{GeV}/c^2/\mathrm{cm}^3$, and the Standard Halo Model \cite{Drukier:1986tm} with a Maxwell-Boltzmann velocity distribution with dispersion velocity $v_0=220\,\,\mathrm{km}/\mathrm{s}$ and escape velocity cut off of $v_{esc}=544\,\,\mathrm{km}/\mathrm{s}$.

The differential elastic DM-nucleus cross section depends on the recoil energy and the DM velocity and is given by
\be
\frac{d\sigma_{SI}}{dE_R} = \frac{\sigma_p\, m_N}{2\, \mu_p^2\, v^2} A^2|F(E_R)|^2 ,
\ee
where $\sigma_p$ is the DM-proton cross section (assumed equal to the DM-neutron cross section). The nuclear form factor $F(E_R)\sim1$ for small momentum transfers, while $A$ is the atomic mass, leading to the coherent enhancement of the cross section.

\subsection{Migdal effect}
\label{sec:migdal}
The rate of ionization due to the Migdal effect for a nuclear recoil energy $E_R$ accompanied by a ionization electron with energy $E_e$ is given by the standard DM-nucleus differential recoil rate in eq. \eqref{eq:dRNRdER} multiplied by the ionization rate \cite{Ibe:2017yqa}
\be
\frac{d^2 R}{dE_R dv} = \frac{d^2 R_{NR}}{dE_R dv}|Z_{\mathrm{ion}}(E_R)|^2,
\label{eq:rate_M}
\ee
where the ionization rate is given by
\be
|Z_{\mathrm{ion}}(E_R)|^2=\frac{1}{2\pi}\sum_{n,\ell}\int dE_e \frac{dp_{q_e}^c(n\ell\to E_e)}{dE_e}.
\ee
Here, $n$ and $\ell$ are the initial quantum numbers of the emitted electron, $q_e=m_e\sqrt{2 E_R/m_N}$ is the electron momentum in the nucleus rest frame immediately after the DM collision, $m_e$ is the electron mass, $m_N$ is the nucleus mass, and $p_{q_e}^c(n\ell\to E_e)$ is the probability to emit an electron with final energy $E_e$. An approximate estimate of the total energy deposited in the detector is given by $E_d = E_e +E_{n\ell}$, where $E_{n\ell}$ is taken to be the binding energy of the $(n,\ell)$ state. This takes into account the fact that the emitted electron may come from an inner orbital and the remaining excited state will release further energy in the form of photons or additional electrons in order to return to the ground state. 

\begin{figure}[t]
\begin{center}
 \includegraphics[width=\textwidth]{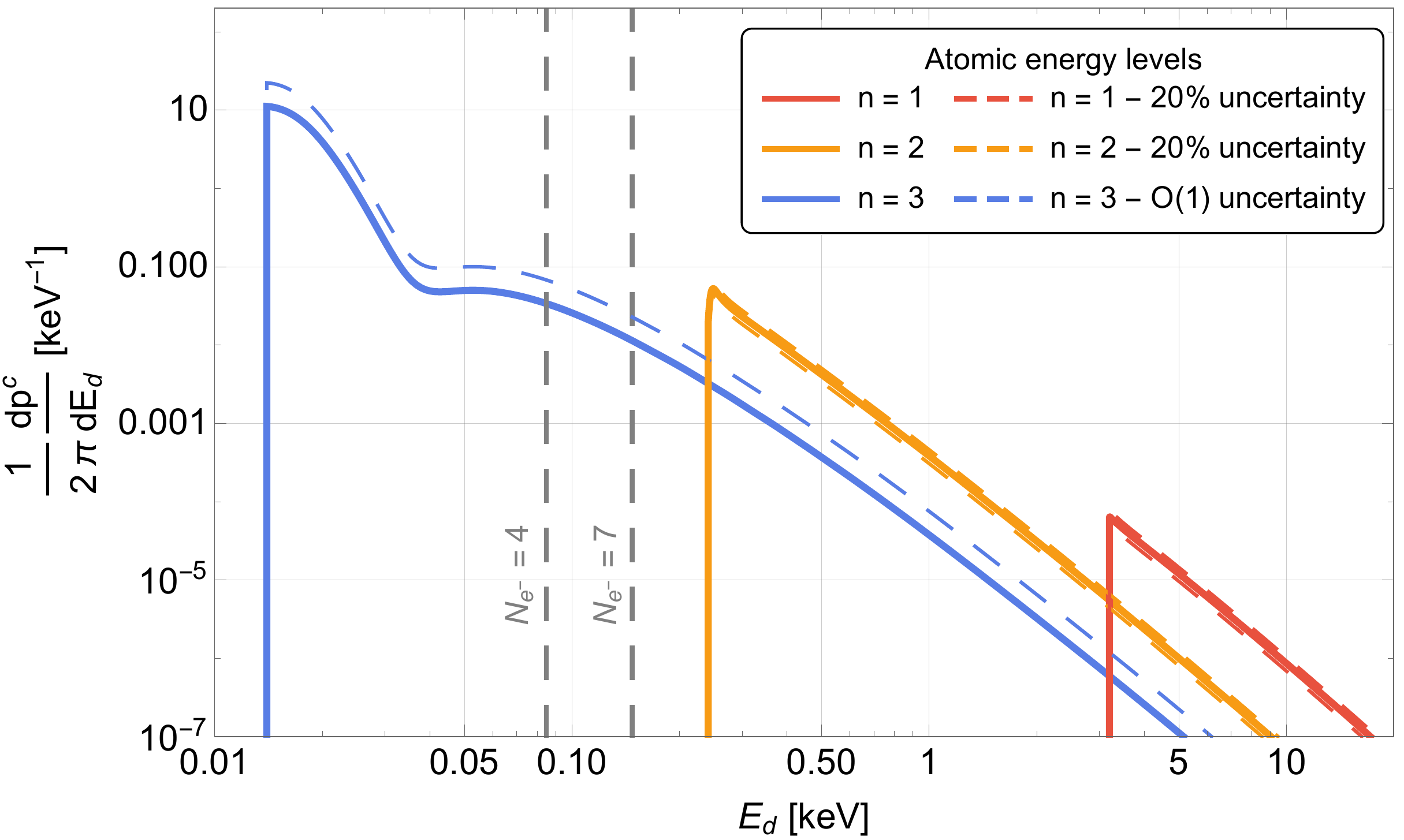}
\caption{Differential ionization probabilities and related uncertainties as a function of the detected energy $E_d$ for isolated argon and different principal quantum number $n$. We show also the 4 and 7 electron thresholds for \DS .}
\label{fig:trans_prob}
\end{center}
\end{figure}

The differential probability rates were computed in Ref. \cite{Ibe:2017yqa} without taking into account the shifts in electronic energy levels because atoms are actually in a liquid (such as in argon or xenon targets) or crystal state (such as in germanium, silicon or sodium iodide detectors). This effect should decrease the ionization energy, and thus, if neglected, should lead to a conservative ionization yield estimate \cite{PhysRevB.13.1649}. Figure \ref{fig:trans_prob} shows the differential ionization probabilities as a function of the detected energy $E_d = E_e +E_{n\ell}$ for isolated argon atoms.  We use the probabilities computed in Ref.~\cite{Ibe:2017yqa}. 

The accuracy of the differential ionization probabilities relies on the fact that the computed wave functions can reproduce the binding energies for the different levels with an accuracy of $\sim\mathcal{O}(20\%)$ and it should provide a correct estimate of the expected signal rate for inner-shell electrons. On the other hand, the prediction for the valence electron shells should be taken as an order of magnitude estimate.\footnote{Private communication with M. Ibe.} However, a new relation between the Migdal process and photo-absorption \cite{Liu:2020pat} gives results comparable with the one obtained by~\cite{Ibe:2017yqa} including also the valence shell.

The different curves of Fig.~\ref{fig:trans_prob} show the contributions for different principal quantum number $n$, where the contributions for different orbital angular momenta in the initial state $\ell$ and all possible final states are summed, and $q_e=1$ eV/$c$. This Figure shows that given the thresholds of 4 or 7 electrons of a hypothetical LAr experiment, the contribution of the valence electrons can maximize the sensitivity to nuclear scattering. This is in contrast with the reported results presented for xenon \cite{Akerib:2018hck,Aprile:2019jmx} and germanium detectors \cite{Armengaud:2019kfj,Liu:2019kzq}, where the detector thresholds are higher than the one needed to see a dominant signal from outer shells.
As a consequence, it would be crucial to have a reliable computation of the transition probabilities in the case of LAr.

We stress that the same considerations discussed here are applicable to neutron scattering and should be taken into account by the experimental collaborations when estimating the radiogenic background contributions. In fact, the original idea was applied to neutron nucleus scattering \cite{Migdal:1941,Landau:1990qp,RUIJGROK1983537,Vegh_1983,Baur_1983}.

In principle, the Migdal effect results can also be compared to DM-electron scattering bounds \cite{Dolan:2017xbu, Baxter:2019pnz, Essig:2019xkx}, in scenarios where the DM couples with equal strength to protons and electrons, as in models with interactions mediated by a dark photon with kinetic mixing \cite{Foot:2004pa, Feng:2009mn}. Such a connection would need some model dependent assumptions that will impact the generality of our results. In addition, making this connection is beyond the scope of this work and as such requires a dedicated study.

\subsection{Photon Bremsstrahlung}

The displacement of the charges of the nucleus and of the electron after the DM-argon scattering leads to photon emission from the polarised argon atom.
Therefore, the elastic nuclear recoil $\chi + N \to \chi + N(E_R)$ is accompanied by the inelastic process $\chi + N \to \chi + N(E_R') +\gamma(\omega)$, where $E_R^{(')}$ is the nuclear recoil energy, while $\omega$ is the photon energy. This process can provide a detectable signal for dark matter masses that produce elastic nuclear recoils below the detector threshold. 

The Bremsstrahlung cross section can be written in terms of the factorised elastic $2 \to 2$ cross section \cite{Kouvaris:2016afs}
\be
\frac{d^2\sigma_{SI}}{dE_R d\omega} = \frac{4 \alpha |f(\omega)|^2}{3 \pi \omega} \frac{E_R}{m_N}\left( \frac{d\sigma_{SI}}{dE_R} \right)\biggl|_{(2\to2)},
\ee
where $E_R$ is the nuclear recoil energy, $\omega$ is the photon energy, $\alpha$ is the fine structure constant, and $m_N$ is the mass of the target nuclei. The atomic scattering factors $f(\omega)=f_1(\omega)~+~i f_2(\omega)$ are tabulated in the NIST Standard Reference Database \cite{doi:10.1063/1.555974}. Notice that for large energies $f_1 \to Z \gg f_2$ (at 4 keV $f_1\sim Z$).
We can then derive the Bremsstrahlung rate as
\be
\frac{d^3 R}{dE_R d\omega dv} = \frac{d^2 R_{NR}}{dE_R dv} \frac{4\alpha |f(\omega)|^2}{3 \pi \omega}\frac{E_R}{m_N}.
\label{eq:rate_BR}
\ee

\subsection{Rates in argon detectors}

We can now show the rates associated with the Migdal effect and photon Bremsstrahlung for dark matter-nucleus scattering in argon detectors. The rates can be obtained by integrating eq.~\eqref{eq:rate_M} and \eqref{eq:rate_BR} over those combinations of $E_R$, $E_e$ (or $\omega$) and $v$ that satisfy momentum and energy conservation. 
In the limit of low momentum transfer both the Migdal effect and the photon Bremsstrahlung process share the same kinematics of inelastic dark matter models \cite{TuckerSmith:2001hy}, where the DM mass splitting $\delta m$ is replaced by the total electronic energy $E_d$ or photon energy $\omega$.  In particular, we have that 
\be
v_{\mathrm{min}} = \sqrt{\frac{m_N E_R}{2 \mu_N^2}} + \frac{\delta}{\sqrt{2 m_N E_R}},
\ee
where $\delta$ correspond to $E_d$ or $\omega$ for the Migdal or Bremsstrahlung processes, respectively. The maximum nuclear and electronic recoil energy for a given DM mass are 
\be
E_{R,\mathrm{max}} = \frac{2 \mu_N^2 v_{\mathrm{max}}^2}{m_N}, \qquad \delta_\mathrm{max} = \frac{\mu_N v_{\mathrm{max}}^2}{2}.
\label{eq:ERmax_deltamax}
\ee
This shows that $\delta_{\mathrm{max}}> E_{R,\mathrm{max}}$ for $m_\chi\ll m_N$ due to the suppression factor $\mu_N/m_N$. Indeed, for $v_{\mathrm{max}}\sim800 \,\mathrm{km/s}\,\sim2.7\cdot10^{-3}\,c$, $m_N\simeq40$ GeV/$c^2$ (the approximate argon mass) and a DM mass of $0.5$ GeV/$c^2$, we find $E_{R,\mathrm{max}} \sim 0.09$ keV, while $\delta_{\mathrm{max}} \sim 1.8$ keV. As a result, there is a range of DM masses for which it is easier to detect the electronic energy originating from the Migdal or the photon Bremsstrahlung processes rather than nuclear recoils, as a consequence of the fact that more energy can be carried off by light or massless particles for a given momentum transfer. 

In particular, eq.~\eqref{eq:ERmax_deltamax} shows that experiments exploiting argon detectors (such as \DS) lose sensitivity for DM masses below $1.8\, {\rm GeV}/c^2$, where $E_{R,\mathrm{max}}\lesssim1$ keV. On the other hand, when considering the Migdal effect or the photon Bremsstrahlung process, the \DS\ experiment is sensitive down to $m_\chi\sim 0.02$ GeV/$c^2$ or $m_\chi\sim 0.04$ GeV/$c^2$ for electron thresholds of $N_{e^-} = 4$ or $N_{e^-} = 7$, respectively.\footnote{Without taking into account the stopping effect of the Earth's atmosphere, mantle and core \cite{Emken:2017qmp, Mahdawi:2017utm, Emken:2018run}, which will be described in Section \ref{sec:Earth_attenuation}.}

\begin{figure}[t]
\begin{center}
 \includegraphics[width=\textwidth]{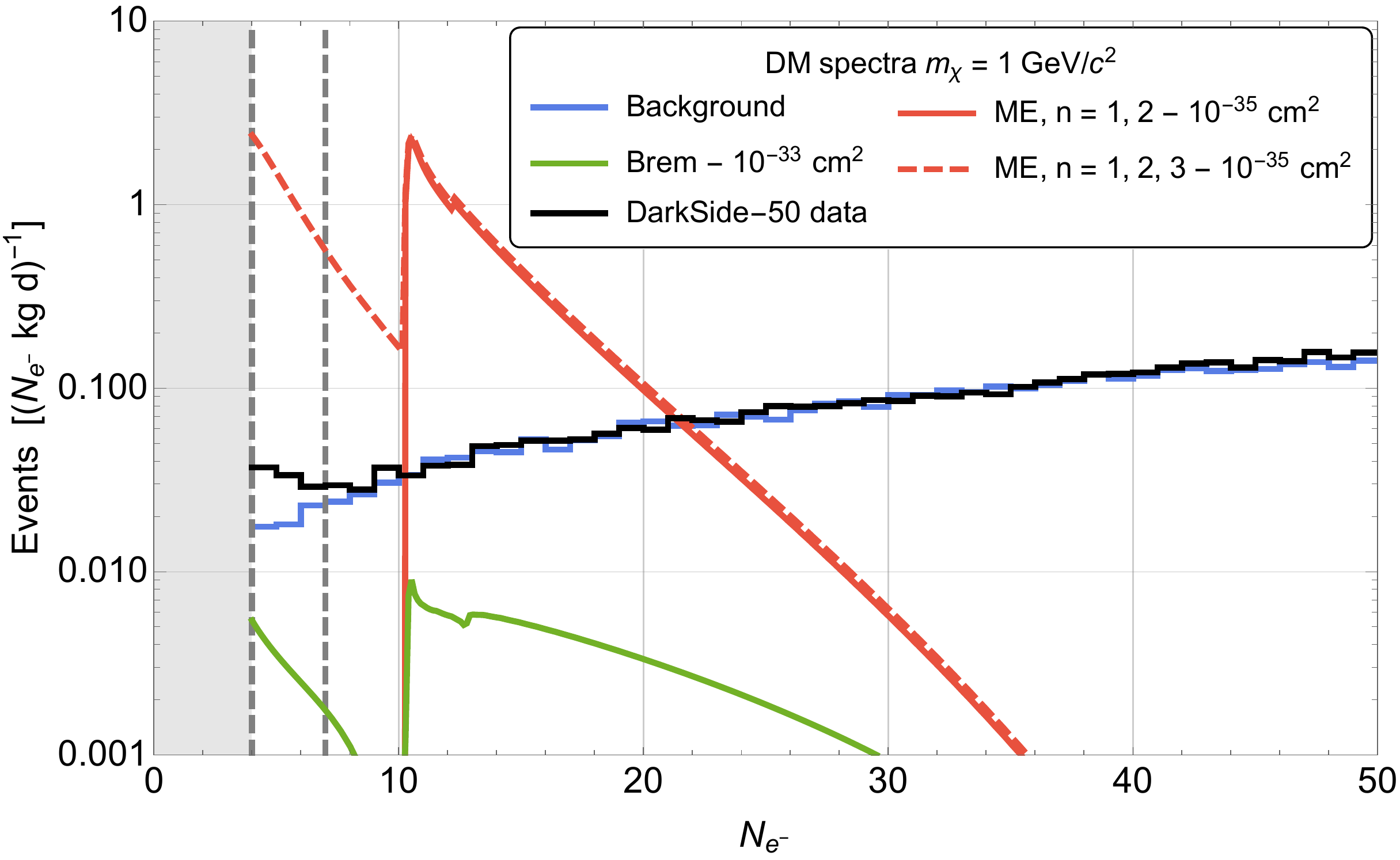}
\caption{Number of events per kg per day for the Migdal effect signal for $m_\chi=1$ GeV/$c^2$ and $\sigma_{\mathrm{SI}} = 10^{-35}$ cm$^2$, for $n=1,2,3$ (red dashed) and $n=1,2$ (solid red).  The green curve shows the number of events for the photon Bremsstrahlung signal for a DM mass of 1 GeV/$c^2$ and $\sigma_{\mathrm{SI}} = 10^{-33}$ cm$^2$. The histograms show the \DS\ spectrum (black) and background  (blue) for an exposure $E= 6786\, {\rm\:kg\:d}$ \cite{Agnes:2018ves}.}
\label{fig:rates}
\end{center}
\end{figure}

In Fig. ~\ref{fig:rates} we show the Migdal and Bremsstrahlung rates as a function of the number of detected electrons, induced by a DM particle scattering on argon with a cross section $\sigma_{SI} = 10^{-35}$ cm$^2$ ($\sigma_{SI} = 10^{-33}$ cm$^2$ for the Bremsstrahlung process) and mass $m_{\chi}=1$ GeV/$c^2$. In order to compute the rates as a function of the number of detected electrons in a LAr detector, we need to transform eq. \eqref{eq:rate_M} and \eqref{eq:rate_BR} with $dE_e/dN_{e^-}$. We find this information using the calibration curve used by the \DS\ collaboration to convert electron recoil spectrum to an average ionization spectrum (Fig. 2 of Ref. \cite{Agnes:2018oej}).\footnote{We assume that a photon produced directly from Bremsstrahlung converts all of its energy into ionization of valence electrons \cite{Kouvaris:2016afs} and that on average an energy of $\mathcal{O}(10)$ eV is needed to produce an ionized electron in argon.}
In this context we neglect the fluctuations associated to the detector response which are discussed in Sec.~\ref{sec:syst}. 
In addition, in the same plot we show the total background (blue histogram) and the measured spectrum (black histogram) from the \DS\ experiment, taken from Fig. 7 of Ref. \cite{Agnes:2018ves}. The figure shows also that the Migdal effect dominates over the Bremsstrahlung rate across all energies. 

The python code we used for the evaluation of the nuclear recoil,  Migdal and Brems\-strahlung rates is publicly available on {\tt GitHub}~\cite{ref:github_DDrates}.

\subsection{Effects of the Earth attenuation}
\label{sec:Earth_attenuation}
Direct detection experiments generally lose sensitivity to strongly interacting DM\footnote{Here we refer to the dark matter-nucleon interaction and not to the DM self-interaction.} because the same interactions that happen in the detector occur also in the Earth atmosphere and crust \cite{Starkman:1990nj,Collar:1992qc,Collar:1993ss,Kouvaris:2014lpa,Kavanagh:2016pyr}. As a consequence, DM particles are slowed down and deflected, reducing the flux of DM particles that arrives at the underground detector. Therefore, there is a critical value of the cross section for which the effect of the Earth attenuation is large enough to make any detector blind to DM interactions. 
The average energy loss for a DM particle traversing the atmosphere or the Earth crust due to elastic scattering is given by \cite{Starkman:1990nj, Kouvaris:2014lpa}
\be
\frac{d\langle E_\chi \rangle}{d x} = -\sum_i n_i(\textbf{r}) \int_0^{E_R^{\mathrm{max}}} dE_R\,E_R\frac{d\sigma}{dE_R},
\ee
where $x$ is the distance traveled, $i$ denotes the different nuclei species encountered and $n_i(\textbf{r})$ is the corresponding number density at position $\textbf{r}$. Given that $E_\chi = m_\chi v^2/2$, a change in the DM energy influences the DM velocity distribution at the target. In particular, assuming that all the DM particles move on a straight line from the atmosphere to the detector, the particle flux must be conserved and one can write the DM velocity distribution at the detector as 
\be
f_{\mathrm{det}}(v_\chi^{\mathrm{fin}}) = e^{2 \kappa d} f(e^{\kappa d} v_\chi^{\mathrm{fin}}),
\ee
where $d$ is the depth of the underground detector and 
\be
\kappa = - \frac{\sigma_p}{m_\chi \mu_p^2}\left(\sum_i n_i(\textbf{r}) \frac{\mu_i^4 A_i^2}{m_i} \right)
\ee
where $\mu_i$ is the reduced mass for the nuclei species $i$, $A_i$ its atomic mass and $m_i$ its nuclear mass. In this equation we have neglected the effect of the form factor $F(E_R) \sim 1$ for light DM.

In order to give an estimate of the Earth's attenuation effect, we compute the DM velocity after the interaction with the Earth's atmosphere and crust using the \texttt{VERNE} code \cite{Kavanagh:2017cru, ref:verne}. We also assume that the DM travel in a straight line from the surface to the detector (minimizing the path through the Earth), in the Laboratori Nazionali del Gran Sasso (LNGS), Italy. For simplicity, we take the maximal DM velocity in the laboratory reference frame, disregarding the daily and annual modulation.\footnote{The value of $v_{\mathrm{max}}$ varies of $\lesssim20\%$ due to the daily and annual modulation (the smaller the cross section, the smaller the effect). This impacts the estimated upper limit of the cross section by a factor of a few, depending on the DM mass (the lower the DM mass, the larger the factor).} More specifically, we compute the maximal DM velocity at the detector depth and set an upper limit on the detector sensitivity on the cross section solving numerically the following equation for $m_\chi$ and $\sigma$
\be
v_\mathrm{max}(m_\chi, \sigma) = \sqrt{\frac{2\,\delta_\mathrm{max}}{\mu_N}}.
\ee
Here the energy $\delta_\mathrm{max}$ is set by the threshold of the experiment. Notice that in this way we overestimate the sensitivity of the experiment to large cross section, but it works as a order of magnitude estimate. We show the results in Section \ref{sec:results}.

\section{Sensitivity calculation and LAr simulated experiment}
\label{sec:inference}
In this work, we adopt a Bayesian approach to infer the upper bound and to estimate the expected experimental sensitivities to the interaction of DM candidates with LAr. Similar approaches for the analysis of DM experimental data have already been deployed ~\cite{Trotta:2006ew,  Roszkowski:2007va, Strege:2012kv, Arina:2013jma, Workgroup:2017lvb,  Liem:2016xpm, Messina:2020pnt, Krishak:2019hlo, Krishak:2019iei}, although they are not frequent among  analyses carried out by the experimental collaborations, as for example~\cite{Abdelhameed:2019hmk,Agnes:2018ves,Agnes:2018oej,Liu:2019kzq,Aprile:2019jmx,Armengaud:2019kfj}.

Within this approach we can compute, at least in principle, the probability of any specific proposition given some state of information. Theoretical models, parameters of interests, and results of experiments before they are carried out are intended as uncertain propositions connected by the rules of probability. Exploiting the Bayes theorem we can update the initial probability for a model or a parameter after new information is available in the form of experimental observations.
The experimental information is fully contained in the so-called likelihood function. This term refers to the conditional probability for the data given the model  when  it is regarded as a function of the model's parameters while keeping the data fixed to the experimental observations.

We use the following notations:
\begin{compactitem}[-]
    \item $D =\{x_i\}$ represents the data,  possibly organised in different classes $i$; 
    \item $E$ is the experimental exposure given in terms of the duration time $T$ of the data-taking period and the fiducial mass $M_{det}$ of the detector. 
    \item $H_r$ represents a specific hypothesis:
    $H_0$ is the background-only hypothesis according to which the known physics processes (backgrounds) are enough to explain the observations;
     $H_{r_S}$ is the background-plus-signal hypothesis for which some DM signal with rate $r_S$ is required to explain the data. We note that the two hypotheses $H_0$ and $H_{r_S}$ are nested since $H_0$ can be obtained for  $H_{r_S}$ by setting $r_S=0$. For what concerns our inferential problem of constraining $r_S$, we will always work within the hypothesis $H_{r_S}$, assuming its validity.
    \item $r_S$ indicates the expected rate of DM interaction for a given $\sigma_{SI}$ per unit mass and time expressed in evt/kg/day. It is also a function of the mass $m_\chi$ of the DM candidate through the cross section.  We take as reference cross section the following values: $\sigma_{SI}({\rm ref.}) =10^{-41}\,, 10^{-37}\,, 10^{-34}\, {\rm cm}^2$ for the nuclear recoil, Migdal electron, photon Bremsstrahlung analyses respectively. 
    \item $r_B$ indicates the rate of the total background events expressed in evt/kg/day;
    \item $\pi(r)$ is the prior probability density function (p.d.f.) for the generic parameter $r$ and encapsulates all the available knowledge on the parameter $r$  before the experiment is carried out;
    \item $\like(r; D)$, or simply $\like(r)$ is the likelihood for the generic parameter $r$ of the hypothesis $H_r$, and coincides with $p(D\, |\, r,\, H_r)$;
    \item $p(r\, |\, D)$ is the posterior p.d.f. for the generic parameter $r$ given the data $D$;
    \item $\boldsymbol{\theta} = (r_B,\dots)$ is the list of parameters necessary to describe the experimental conditions or theoretical assumptions which are not exactly known but can vary according to they prior p.d.f. $\pi(\boldsymbol{\theta})$, these are the so-called nuisance parameters as we are not explicitly interested in inferring their posterior values;
    \item $\Omega$ is the nuisance parameters space.
\end{compactitem}

We recall that the posterior p.d.f for the parameters of the model can be computed by means of the Bayes theorem as:
\begin{equation}
    p(r_S,\, \boldsymbol{\theta}\, |\, \{x_i\},\, H_{r_S}) = 
    \frac{ p( \{x_i\}\, |\, r_S,\, \boldsymbol{\theta},\,  H_{r_S}) \pi(r_S,\, \boldsymbol{\theta}\, |\,H_{r_S})}
    {\int_\Omega {\int_0^{\infty} {p( \{x_i\}\, |\, r_S,\, \boldsymbol{\theta},\,  H_1)\pi(r_S,\, \boldsymbol{\theta} |H_1) dr_S} d\boldsymbol{\theta}}},
    \label{eq:bayes}
\end{equation}
with 
\begin{equation}
    \like(r_s,\, \boldsymbol{\theta}) \equiv p( \{x_i\}\, |\, r_S,\, \boldsymbol{\theta},\,  H_{r_S}).
\end{equation}
The marginal p.d.f. of the parameter of interest $r_S$ is given by:
\begin{equation}
    p(r_S\, |\, \{x_i\},\,  H_{r_S}) = 
    \int_{\Omega}{p(r_S,\, \boldsymbol{\theta}\, |\, \{x_i\},\, H_{r_S}) d\boldsymbol{\theta}},
\end{equation}
and similarly for any other parameter of the model.

\subsection{Upper bounds and experimental sensitivity}
\label{sec:upper_bounds_exp_sensitivity}

\subsubsection{90\% Credible Interval upper bound}
We compute the upper bound for the DM signal as the 90\% Credible Interval (C.I.). This is defined as the value of $\sigma_{SI}(m_\chi)$ corresponding to the 90\% quantile of the posterior p.d.f. for $r_S$:
\begin{equation}
    r_S(90\%\, {\rm C.I.}): \\
    \int_0^{ r_S(90\%\, {\rm C.I.})} p(r_S\, |\, \{x_i\},\, H_{r_S})\, dr_S = 0.9.
\label{eq:quantile_def}
\end{equation}
In the Bayesian approach the upper bound is a statement on the true value of the parameter of interest. 
The quantity $r_S(90\%\, {\rm C.I.})$ has to be interpreted as the value below which we believe at 90\% probability level the true value of $r_S$ lies, given the present experimental information. 

\subsubsection{Prior choice}
It is evident from eq.~\eqref{eq:bayes} that the posterior p.d.f. depends on the priors on all parameters. However, we have to distinguish  the effect due to the priors on  nuisance parameters from the one due to the prior on the parameter of interest. The former has the effect of averaging the posterior over the nuisance parameters space, that is an elegant way of propagating systematic effects on the parameter of interest. In addition, the prior of nuisance parameters is often a parametrization  of calibration measurements. The latter, although indispensable to invert the probability and get the posterior, has a degree of `subjectivity' with potentially a significant impact on the posterior. The prior $\pi(r_S)$ represents the knowledge on $r_S$ before the experiment is carried out, and gets updated by a factor proportional to the likelihood of the observed data. It is a critical term in many respects, and it should reflect the researcher state of knowledge. Especially for searches where the sought quantity is unknown and the search is pushed to the limit of the experimental sensitivity, the input from $\pi(r_S)$ might have sizable effect on the posterior. The prior has thus to be well justified and the posterior sensitivity to different prior choices needs to be explored.

In our case, $r_S$ depends on $m_\chi$ and on $\sigma_{SI}$. For $m_\chi$ we chose a flat prior.
For $\sigma_{SI}$, to explore the sensitivity of the upper bound to the prior choice, we studied its behaviour for a Migdal-only signal at $m\chi = 1$~GeV/$c^2$ and with $n= 1,2$. We generated a pseudo-dataset from the background template and performed a fit using different priors. We tested, in two possible domain ranges, namely $D_1 = [10^{-40}, 10^{-37}]\:{\rm cm^2}$ and $D_2 = [10^{-44}, 10^{-37}]\:{\rm cm}^2$, four different prior choices: a uniform prior, a wider gamma prior with a shape $k = 1$ and a scale $\theta = 10^{-37}\:{\rm cm}^2$, a narrower gamma prior with a shape $k = 1$ and a scale $\theta= 10^{-38}\:{\rm cm}^2$ and a uniform prior in  $\log(\sigma_{SI}/{\rm cm^2})$ (we will call this prior ``loguniform''). In principle, there is no need to restrict the domain to a finite interval, but both the uniform and the loguniform distributions are not normalizable otherwise. In addition, there could be physical motivations that define a reasonable interval. In our opinion, for our problem, from above the natural constraints come from upper bounds imposed by previous experiments as for example Xenon1T~\cite{Aprile:2019jmx} or CRESST-III~\cite{Abdelhameed:2019hmk}, which for the chosen configuration exclude at $90\%$ C.L. cross sections of the order $10^{-38}\:{\rm cm}^2$.
From below we can use two arguments: the first is that below $10^{-44}\:{\rm cm}^2$ the rate would be dominated by neutrino coherent scattering \cite{PhysRevD.90.083510} (the so-called `neutrino floor'), the second is that the experimental sensitivity  does not extend below $10^{-40}\:{\rm cm}^2$, and will be discussed in Sec.~\ref{subsec:sensitivity}. Therefore the choice of the domain $D_1$ is driven by physical considerations about a LAr experiment with features similar to \DS\, while the choice of the domain $D_2$ extents up to the maximum experimental sensitivity that an experiment of this kind can reach before hitting the neutrino floor.

\begin{figure}[t!]
    \centering
    \includegraphics[width = .48\textwidth]{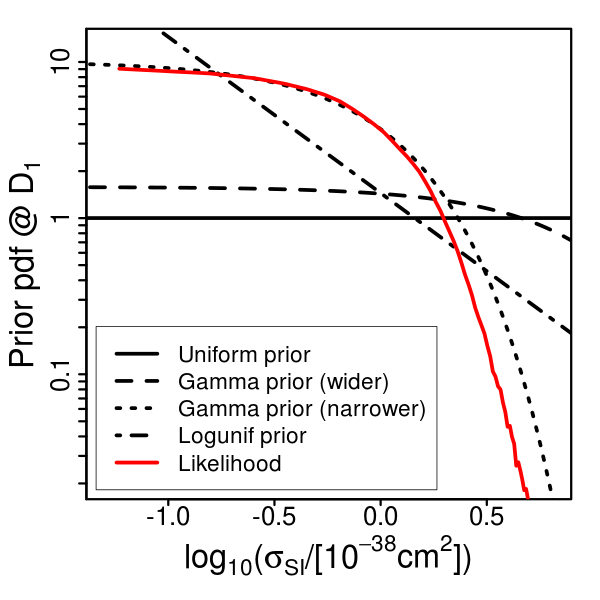}
    \includegraphics[width = .48\textwidth]{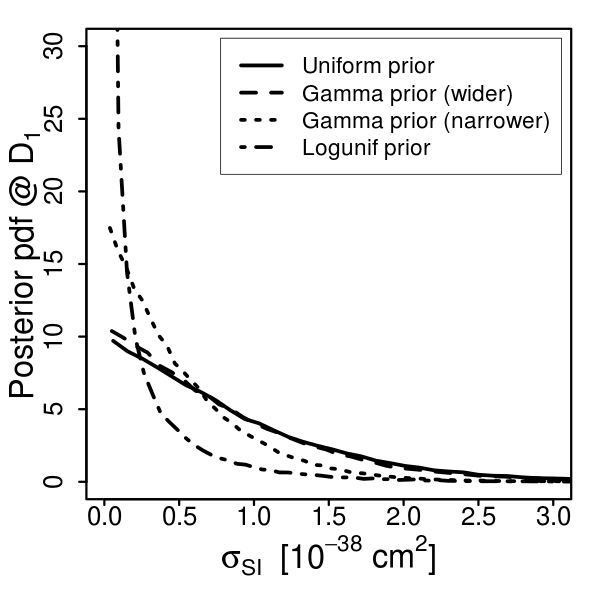}
    \caption{{\bf Left}: possible prior choices for the parameter $\sigma_{SI}$. {\bf Right}: resulting posterior p.d.f. after a fit over a pseudo-dataset generated from the background template.}
    \label{fig:priorcomparison}
\end{figure}

The results, in terms of posterior p.d.f. for $r_S$ and $\sigma_{SI}$ are reported in Figure~\ref{fig:priorcomparison}: here we show, as an example, all p.d.f. in the $D_1$ domain (however the posterior p.d.f. using the domain $D_2$ are very similar to the one showed in Figure~\ref{fig:priorcomparison} and the differences in terms of the $90\%$ C.I. upper bounds are reported in Table~\ref{tab:prior_quantiles}). There is no much difference between the uniform and the wider gamma cases, and that is because, as one can see from the left part of Figure~\ref{fig:priorcomparison}, these priors are quite flat in the region where the likelihood (the red line in the plot) is mostly informative; on the other side, for the loguniform and the narrower gamma, this is not true in the range where the experiment sensitivity is lost (below $~10^{-39}\:{\rm cm}^2$), and this is reflected both in the posterior p.d.f. and the $90\%$ C.I. upper bound, as reported in Table~\ref{tab:prior_quantiles}. We can therefore quantify the dependence of the bound from the prior choice in a factor as big as $10$, and this confirms the importance of choosing the prior in a reasonable and coherent way. For simplicity and for reason that would be clear in the Sec.~\ref{subsec:sensitivity} in the rest of the paper we will report upper bounds obtained using a flat prior.

%%%%%%%%%%%%%%%%%%
\begin{table*}[!ht]
\def\arraystretch{1.1}
\centering
\begin{tabular}{|c||c|c|c|c|}
\hline
Prior p.d.f. & Uniform & Gamma [w] & Gamma [n] & Loguniform \\
\hline
$\sigma_{SI}(90\%{\rm C.I.})$ $[ 10^{-38} \: cm^2 ] \:\:@ \:D_1$ & $ 1.36 $  &  $ 1.32 $  &  $ 0.96 $ & $ 0.54 $ \\
\hline
$\sigma_{SI}(90\%{\rm C.I.})$ $[ 10^{-38} \: cm^2 ] \:\:@ \:D_2$ & $ 1.36 $  &  $ 1.30 $  &  $ 0.95 $ & $ 0.17 $ \\
\hline
\end{tabular}
\caption{$\sigma_{SI}(90\% C.I.)$ for each of the prior choices. The [w] index states for ``wider'' and the [n] index stands for ``narrower''.}
\label{tab:prior_quantiles} 
\end{table*}
%%%%%%%%%%%%%%%%%%%%%%%%%%%%%%%%%%%%%%%%%%%%%%

\subsubsection{Experimental sensitivity  and Bayes factor}
\label{subsec:sensitivity}
A meaningful way to report the experimental sensitivity to the sought phenomenon which is as much as possible independent form the priors is given by the Bayes factor. 

The posterior p.d.f. for a signal rate of $r_S$ given a background rate of $r_B$, and $x$ observed events can be normalized to the posterior for $r_S=0$, obtaining:
\begin{equation}
    \frac{ p(r_S\, |\, x,\, r_B ) }{p(r_S=0\, |\, x,\, r_B )} = \frac{\like(r_S \,|\, r_B)}{\like(r_S=0 \,|\, r_B)}\cdot 
    \frac{\pi(r_S)}{\pi(r_S=0)}
\end{equation}
where the first factor on the right hand side is called Bayes factor. It is independent on priors and it is simply given by the likelihood ratio of the two hypotheses:
\begin{equation}
    \mathcal{R}(r_S\, |\, x,\, r_B ) = \frac{\like(r_S \,|\, r_B)}{\like(r_S=0 \,|\, r_B)}.
\end{equation}
The properties  of the $\mathcal{R}$ function have  been discussed in great detail for a similar  case study in ref.~\cite{Astone:1999wp}.
Here we only mention that $\mathcal{R}$  has the  probabilistic interpretation of  hypotheses belief updating ratio. It is equal to 1 in the limit $r_S \to 0$, in this limit the experimental sensitivity is lost and thus the experiment does not change the relative belief. While $\mathcal{R}\to 0$ for large $r_S$, where the posterior density for $r_S$ vanishes no matter how strong it was before.
In addition, the quantity $\mathcal{R}$ is used as test statistic in the frequentist approach to limit settings, for details see Ref.~\cite{PhysRevD.98.030001} (section 39-Statistics) and Ref.~\cite{Cowan:2010js, Algeri:2020pql}.

In the simple case of a Poisson process of intensity $(r_S+r_B)E$, where $E$ is the exposure, and observed counts $x$ the likelihood is proportional to $e^{-(r_S+r_B)E}\, \left[(r_S+r_B)E\right]^x$, thus:
\begin{equation}
    \mathcal{R}(r_S\, |\, x,\, E,\, r_B ) =  e^{-r_SE}\left(1+ \frac{r_S}{r_B} \right)^x
\end{equation}

\begin{figure}[t!]
    \centering
    \includegraphics[width = .49\textwidth]{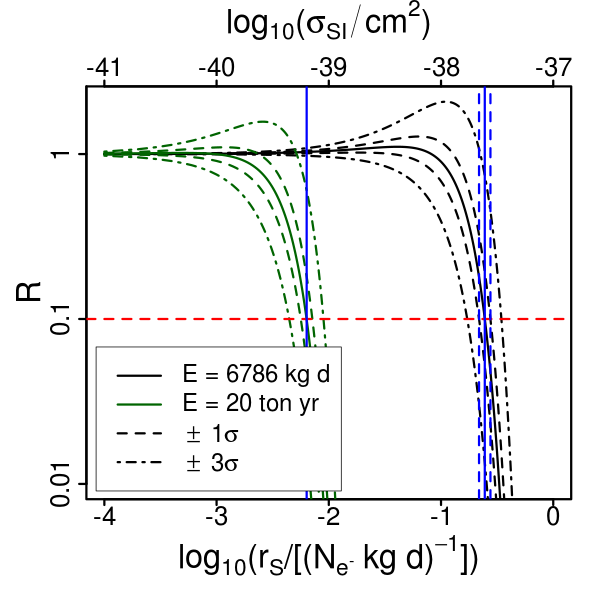}   \includegraphics[width = .49\textwidth]{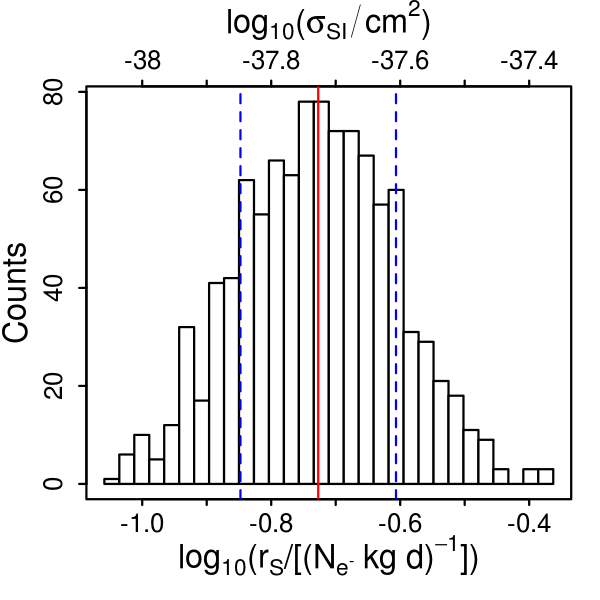}
    \caption{{\bf Left}: Bayes factor for a Migdal-only signal at $m_\chi = 1$~GeV/$c^2$ and $n= 1,2$ obtained using pseudo-datasets generated from the background-only likelihood with the current \DS\ exposure  $E=6786\:{\rm kg\:d}$ (black lines) and an exposure $E=20\:{\rm ton\:y}$ (green lines), which is $\sim10^3$ times the current exposure. The dashed lines and the dashed-dotted lines represent the  Bayes factor computed varying the expected rate by $\pm \: \sigma$ and $\pm 3 \: \sigma$, respectively. {\bf Right}: Histograms of $r_S(90\%\, {\rm C.I.})$ for a Migdal-only signal at $m_\chi = 1$~GeV/$c^2$ and $n= 1,2$ obtained using pseudo-datasets generated from the background-only likelihood with an exposure ${E=6786\:{\rm kg\:d}}$.}
    \label{fig:R_factor}
\end{figure}

We evaluated the $\mathcal{R}$ function for the same configuration used in the previous section. To explore how $\mathcal{R}$  changes when data differ from the expectation (or pseudo-data in this case) due to statistical fluctuations, we consider two additional pseudo-dataset obtained from the previous one by letting pseudo-data fluctuate by $\pm$ 1 (Poisson) standard deviation. The results are reported in the left side of Figure~\ref{fig:R_factor} (black lines), where the green lines represent the corresponding results for a greater value of the exposure (i.e. $20{\rm\:ton\:yr}$ which is roughly $10^3$ times the current exposure $E=6786{\rm\:kg\:d}$).
From this figure we see that with the current exposure the informative region where the experiment has sensitivity and  $\mathcal{R}\to0$ is starting from $r_S\sim 10^{-39}$~cm$^2$. Any conventional value of $r_S$ in this region would be representative of the experimental sensitivity. The statistical uncertainty associated to this value is computed using pseudo-data generated varying the expected rate by $\pm\sigma$. A possible value of $r_S$ representative of this region could be such as that $\mathcal{R}(r_S) = 0.10$, which corresponds to a probability update ratio of 10\% with respect to the null hypothesis. 

As well described in ref.~\cite{Astone:1999wp}, in order to extract any probabilistic statement on $r_S$ from $\mathcal{R}$ one has to add the information about the prior.
We would like to stress that there's a conceptual difference in using the $90\%$~C.I. upper bound or taking the $r_S$ such that $\mathcal{R} = 0.1$: the former is the cumulative of the posterior p.d.f. and then it takes into account all the possible values of $r_S$ from $0$ to $r_S(90\%\, {\rm C.I.})$ as well as the prior choice; the latter is the likelihood ratio and it is a punctual comparison, namely it takes into account only one single possible value of $r_S$ (the one which solves $\mathcal{R}(r_S) = 0.1$), and it is prior independent.

In Table~\ref{tab:Rvsquantile} we show how the two methods gives very similar results. For the $r_S(90\%\, {\rm C.I.})$ method, the results reported in this table are obtained using the procedures described in the next subsection. We also show the results of the projection for different background rates and to higher exposures, reproducing the expected $1/\sqrt{RE}$ scaling, with $RE = E/E_0$ and $E_0 = 6786{\rm\:kg\:d}$.

%%%%%%%%%%%%%%%%%%
\begin{table*}[!ht]
\def\arraystretch{1.1}
\centering
\begin{tabular}{|l|c||c|c|}
%$\log{}_{10} (\sigma_{SI}(90\%{\rm C.I.}) / cm^2 )$
\hline
   & & \multicolumn{2}{c|}{ $\log{}_{10} (\sigma_{SI}(90\%{\rm C.I.}) / cm^2 )$} \\
\hline
$E_0 = 6786{\rm\:kg\:d}$ & $r_{B_0} = 3.64{\rm\:(N_{e^-}\:kg\:d)^{-1}}$& Using & Using $r_S(90\% {\rm C.I.})$ \\
& &$\mathcal{R}(r_S) = 0.1$ & defined by eq.~\eqref{eq:quantile_def} \\
\hline
$E = E_0$ & $r_B = r_{B_0}$ & $ -37.68 \pm 0.19 $  &  $ -37.73 \pm 0.12
 $ \\
\hline
$E = 20  E_0$ & $r_B = r_{B_0}$ &$-38.37 \pm 0.18$ & $-38.15 \pm 0.09$ \\
\hline
$E = 100 E_0$ & $r_B = r_{B_0}$ &$-38.71 \pm 0.19$ &  $-38.50 \pm 0.09$ \\ 
\hline
$E = 1076 E_0$& $r_B = r_{B_0}$ &$-39.20 \pm 0.19$ &  $-39.15 \pm 0.11$ \\
\hline
$E = E_0$& $r_B = r_{B_0}/10$ &$ -38.18 \pm 0.19$ &  $-38.21 \pm 0.14$ \\
\hline
$E = E_0$& $r_B = r_{B_0}/10^2$ &$-38.63 \pm 0.20$ & $-38.66 \pm 0.15$  \\
\hline
\end{tabular}
 \caption{Upper bound results in two different methods for a Migdal-only signal at $m_\chi = 1$~GeV/$c^2$ and $n= 1,2$ using pseudo-datasets generated from the background-only likelihood.}
\label{tab:Rvsquantile} 
\end{table*}
%%%%%%%%%%%%%%%%%%%%%%%%%%%%%%%%%%%%%%%%%%%%%%

\subsubsection{Expected sensitivity}
The expected sensitivity can be quantified either with the upper bound or with the $\mathcal{R}$ function using pseudo-data generated with the likelihood under the background-only hypothesis. In the following of this work we quote only upper bounds at $90\%\, {\rm C.I.}$ obtained with a flat prior on $r_S$. This choice is motivated by the pragmatic argument that a bayesian limit is easily comparable with results obtained with different statistical strategies. This approach allows also a straightforward procedure to fold in different priors. 
Pseudo-data allow to explore the sensitivity to a given signal rate for a specific exposure. For the same exposure, we can generate pseudo-data including also the statistical fluctuation of the observed number of events allowing to study the dependence of the sensitivity to the expected statistical fluctuation if no signal were present. In principle with the same technique we could also explore the impact of systematic effects.

To compute the expected sensitivity and produce the results reported in the right column of Table~\ref{tab:Rvsquantile}, we generated 1024 pseudo-experiments each of which consisting of a set of $D=\{x_i\}$ pseudo-data simulated from the likelihood under the background-only hypothesis. We then compute the posterior p.d.f. and the upper bound $r_S(90\%\, {\rm C.I.})$. As central value for $r_S(90\%\, {\rm C.I.})$ we take the sample mean, and as uncertainty the sample standard deviation.
An example of the histogram of $r_S(90\%\, {\rm C.I.})$ is given in the right side of fig.~\ref{fig:R_factor}.

\subsection{The {\sc tea-lab} simulated LAr experiment}\label{sec:tealab}
In this work we are interested in studying the impact of the Migdal electron and photon Bremsstrahlung to the sensitivity of LAr experiments to light DM. We expect sizable effects for DM masses below $\sim 2\,{\rm GeV}$, where the sensitivity of current experiments is lost.

We develop a toy simulation of a LAr experiment that we call {\sc tea-lab} (Toy Experiment Analysis of Liquid Argon Behaviour) loosely inspired by \DS. 
\DS\ is a Liquid Argon Time Projection Chamber (LAr TPC) ~\cite{Agnes:2014bvk} operated in the LNGS in Italy. 
The LAr TPC is red-out by Photomultipliers (PMTs) sensitive to the scintillation light produced by the 
ionizing events in the active LAr target, the so-called `S1' signal. The ionization electrons produced at this stage, and surviving the recombination process, are drifted by the TPC electric field to the liquid surface, where they are extracted into an argon gas layer. The electric field in the gas is large enough to accelerate the electrons which excite the argon such to generate a secondary scintillation signal, `S2'. 
The lowest threshold is achieved by exploiting the high gain of the S2 signal and corresponds to a number of primary ionization electrons  $N_{e^-}=4$. This result has been achieved by the DarkSide collaboration thanks to a detailed understanding and calibration of the detector response~\cite{Agnes:2018ves, Agnes_2017, Agnes:2017grb, PhysRevD.88.092006, PhysRevD.91.092007, Agnes:2018fwg}. 
The \DS\ spectra, as given in Fig.~7 of Ref.~\cite{Agnes:2018ves}, refer to S2-only events and correspond to a $6786.0\, \rm{kg\, d}$ exposure (corrected for the fiducialization cut). For a detected energy
 $E_d>0.05\,\rm{keV}_{ee}$, well below the analysis threshold, the LAr TPC is fully efficient \cite{Agnes:2018oej}, thus no efficiency correction is needed.
 
{\sc tea-lab} assumes the \DS\ total background spectra and includes some relevant experimental effects as described below.

\subsubsection{The likelihood function}
The likelihood function represents the connection between the parameter of the model, both theoretical and experimental, with the observed data $D$.
The experimental observable is the number of nuclear recoils $x_i$ that occur producing $i$ primary ionization electrons. 
We factorise the likelihood\footnote{This implies to assume  there is no effect that connects the background and signal yields. This is quite a strong assumption, however we believe it is appropriate for our simulated experiment. The likelihood can be generalised adding a nuisance parameter common to $\mathcal{L}_S$ and $\mathcal{L}_B$ with an appropriate prior.}
 in three terms:
\begin{equation}
    \like = \like_C \times \like_B \times \like_S
\end{equation}
with the following meaning:
\begin{compactitem}[-]
 \item $\like_C$: describes the probability that in each observation bin $i$ the number of counts is $x_i$. This is assumed Poisson distributed given the expected number of counts $\lambda_i$. The counts in different bins are taken as independent. Under these assumptions:
\begin{equation}
    \like_C \left(r_S,\, r_B,\, \boldsymbol{\theta}; \{x_i\}\right) \equiv
    p(\{x_i\}\,|\,r_S, r_B, \boldsymbol{\theta}, H) =
    \prod_{i=1}^{N_{bin}} \frac{\lambda_i^{x_i}}{x_i!}\, e^{-\lambda_i},
\end{equation}
with:
\begin{equation}
   \lambda_i = E \, [r_S\, S_i + r_B\, (B_i +  LowNe_i )]
\end{equation}
where $E=T M_{det}$ is the experimental exposure.
The quantities $S_i,\, B_i$, and $LowNe_i$  are associated respectively to the DM signal, to the total background, and to a possible background source at low $N_{e^-}$ to account for the excess of events, assumed to be due to some not completely understood experimental effect \cite{Agnes:2018ves,Agnes:2018oej},  visible in the published spectrum below $N_{e^-}=7$.

\item $\like_B$: describes how the background template is affected by systematic uncertainties. We don't have the information about how to include the different systematic effects, and a thorough implementation goes beyond the scope of this paper. %However, 
%we decided to include bin dependent Gaussian uncertainties 
%with a standard deviation compatible with a  simulated sample size of $10^5$ events: this corresponds to a $3$-$6\%$ uncertainty for $N_{e^-} \geq 10$.
However, we decided to include bin dependent Gaussian uncertainties with a standard deviation in the range of 3-6\% for $N_{e^-} \ge 10$ to account for the statistical fluctuation of the background model as given in fig.\,7 Ref.\cite{Agnes:2018ves}.
We don't account for any systematic effect on the background spectra.
For the  $LowNe$ background we proceeded differently. We parametrized its contribution  in the range $4 \le N_{e^-}\le 7$ with a 2-parameter function, and use the parameters to control its contribution ranging from no contribution at all to something similar to what is visible in the \DS\ spectra. 
This likelihood term is given by
\begin{equation}
    \like_B = 
    \prod_{\{bkgd\}}\prod_{i=1}^{Nbin}\mathcal{N}(\mu=bkgd_i, \sigma=\sigma_{bkgd_i})
\end{equation}
with ${bkgd} = \{B, LowNe(p_0,\, p_1) \}$.
For the $LowNe$ we introduced explicitly here the dependence on the two parameters used to model it. Here, these parameters are assumed as given, in the next section we discuss how we deal with their uncertainties.

\item $\like_S$: this factor depends on the systematic uncertainties on the signal template $S_i$. We are considering 2 effects: one that parametrises the uncertainty in the emission probabilities as discussed in section~\ref{sec:syst}, and the other that describes the experimental efficiency to convert the energy of the Migdal electron in primary ionization electrons. $\like_S$ is given by
\begin{equation}
   \like _S = 
    \delta [ S_i - S_i(\mathbf{f}, N_{e^-}^{max}, \epsilon)] 
\end{equation}
where $S_i(\mathbf{f}, N_{e^-}^{max}, \epsilon)$ represents the probability of the signal to give rise to an event. It depends on the maximum number of ionization electrons that can be in principle produced $N_{e^-}^{max}$ and on the efficiency $\epsilon$ of the production mechanism and detection as it is explained in section ~\ref{sec:syst}.  Finally, the parameter $\mathbf{f}=(f, f_{val})$ controls the contribution of the emission probability due to the inner ($f$) and valence ($f_{val}$) shell and are used to parametrise the systematic uncertainties associated to their calculation. The central value of the calculation is obtained for $(f, f_{val})=(1,1)$.  Independently of $N_{e^-}$ and $\epsilon$, the contribution of the different shells can be expressed as:
\begin{equation}
S_i(f) = f\, s_i(n=1,2) + f_{val}\, s_i(n=3) 
\label{eq:signal_components}
\end{equation}
where $n$ indicates the electron shell(s) considered. 
\end{compactitem}

\subsubsection{Simplified treatment of systematic effects}
\label{sec:syst}
The likelihood function described above is quite general and can be used to parametrise several systematic effects. However, any complete
description of such effects requires a detailed knowledge of the detector which is beyond the scope of this work. For this reason, we leave the description of the systematic effects to the experimental collaboration except for a simplistic treatment of the following few relevant effects.
\begin{compactitem}[-]
\item {\bf Background rate normalization}. The total background rate is controlled by the nuisance parameter $r_B$. Although it can be predicted by an accurate simulation, we leave it float with a uniform prior and then constrain it to few percent (see Fig.~\ref{fig:correlation}) in the fit with high $N_{e^-}$ spectrum. 

\item {\bf Low $N_{e^-}$ excess}.  The most conservative way to deal with this not completely understood effect is to remove from our fit the region where effect emerges. By setting a threshold $N_{e^-}=7$, this region is removed. By lowering the threshold to $N_{e^-}=4$, the $LowNe$ contribution becomes important. For this configuration we explored 2 options. The first, and more conservative option, is to let the fit account for the  excess with a signal contribution, and thus weaken the limit. The second is to model this effect with a 2-parameter function and assign it to an unknown background contribution. 
We decided to report results also with this configuration as it gives the level of sensitivity one may reach if the excess were understood. In this case we assign normal probabilities $\pi(p_0)$ and $\pi(p_1)$ to the parameters as given by the fit and then we marginalise these parameters in the limit computation.

\item {\bf Contribution of the electron shells}: The contribution of the outermost electron shell to the Migdal effect in LAr is affected by large theoretical uncertainties, and the result given in \cite{Ibe:2017yqa} can safely be taken as an order of magnitude estimate (see Sec.~\ref{sec:migdal}). For this reason we decided to explore its impact by setting $f_{val}=0,2$ in eq.~\eqref{eq:signal_components}. These values correspond to a variation of $\pm$100\% around the estimated contribution of the valence electron give by $f_{val}=1$. The additional configuration 
explored, is to consider $f_{val}$ as a nuisance parameter with a flat prior $\pi(f_{val})$ in the range [0,2] and marginalize this parameter in the limit evaluation.
For the inner shells ($n=1,2$) we included a gaussian prior $\pi_{f} = \mathcal{G}({\rm mean}=1, {\rm std}=0.2)$ to account for the $\mathcal{O}(20\%)$ theoretical uncertainty in their calculation \cite{Ibe:2017yqa}.

\item {\bf Fluctuation induced by the detector response}:
The conversion of the Migdal electron energy ($E_e$) into a number of ionization electrons ($N_{e^-}$) is a stochastic process which depends on the details of the liquid Ar ionization and excitation processes and on the detector response\footnote{We don't introduce  explicitly any detector resolution effect. This has a small effect on the sensitivity studies we are carrying out since, as it is shown in Sec.~\ref{sec:results}, also the binomial fluctuations don't affect our conclusions. We also checked that doubling the resolution the limit doesn't change significantly since it is fairly insensitive to the signal shape. Clearly, for any experimental analysis the detector resolution is an important ingredient.}. To model these fluctuations we proceed with a very crude approximation of what is done in Ref. \cite{Agnes:2018oej,Agnes:2018ves, Agnes:2018mvl}. The average number of ionization electron $\langle N_{e^-}\rangle$ per keV is taken from Fig. 2 of Ref.~\cite{Agnes:2018oej}, while the maximum number of ionization electron can be estimated as  $N_{e^-}^{max} = E_e/W$, where $W \simeq 19.5$~eV \cite{Doke:2002oab, Agnes:2017grb, Kimura:2020bxf} is the effective LAr ionization work function required to produce an electron-ion pair. The final efficiency can be estimated as $\epsilon =\langle N_{e^-}\rangle/N_{e^-}^{max}$. Thus the probability of having a certain number $N_{e^-}$ of ionization electrons to produce the detected signal is given by the binomial distribution:
\begin{equation}
    P(N_{e^-}, E_e) = \mathcal{B}(N_{e^-} \,|\, p = \epsilon, n = N_{e^-}^{max}). 
\end{equation}
As a reference the width for $N_{e^-}=10(30)$ is $\sigma =1.2(3.8)$. 
This probability depends on the energy $E_e$, and thus it is convoluted with the energy spectrum of the Migdal electron emission.
\end{compactitem}

\subsection{Analysis model implementation}
The computation of the posterior p.d.f., even for models relatively simple as those described in the previous section, is often only possible by Monte Carlo integration. The most common way to solve problems of this kind is by sampling the not normalised posterior distribution by a Markov Chain Monte Carlo (MCMC).
For our study we used the general analysis framework {\tt R} \cite{ref:R} and the MCMC algorithm called {\it Gibbs Sampler} as implemented in  {\tt JAGS} \cite{Plummer03jags:a} and interfaced with {\tt R} in the package {\tt rjags} \cite{ref:rjags}. The details of the implementation and the source code for the  analysis are publicly available on {\tt GitHub}~\cite{ref:github_LAr-MigdalLimits}.
The Monte Carlo simulation gives the unnormalised posterior p.d.f. of the parameters of interest sampled using the Gibbs algorithm. The results reported in this work are obtained with a single Markov chain with $10^5$ steps, which is enough to guarantee the chain thermalization.

\section{Sensitivity to Migdal electron and photon Bremsstrahlung}
\label{sec:results}

Having computed the rates for the Migdal effect and the photon Bremsstrahlung process, and described the {\sc tea-lab} simulation, we are able to study the expected sensitivity to low mass dark matter using {\sc tea-lab} as a case study.
We assume no isospin violation (the neutron and proton cross sections are equal). The nuclear recoil contribution was ignored in the Migdal and Bremsstrahlung signal models because it is small if compared with the electron recoils, for masses below $1.8$ GeV/$c^2$.
For our simulated {\sc tea-lab} experiment we generated, as already explained in Sec.~\ref{sec:upper_bounds_exp_sensitivity} and~\ref{sec:tealab}, a dataset from the background-only template, including this time the $LowNe$ excess. We take always as a reference exposure $E = 6786{\rm\:kg\:d}$.

\begin{figure}[!t]
\begin{center}
 \includegraphics[width=.95\textwidth]{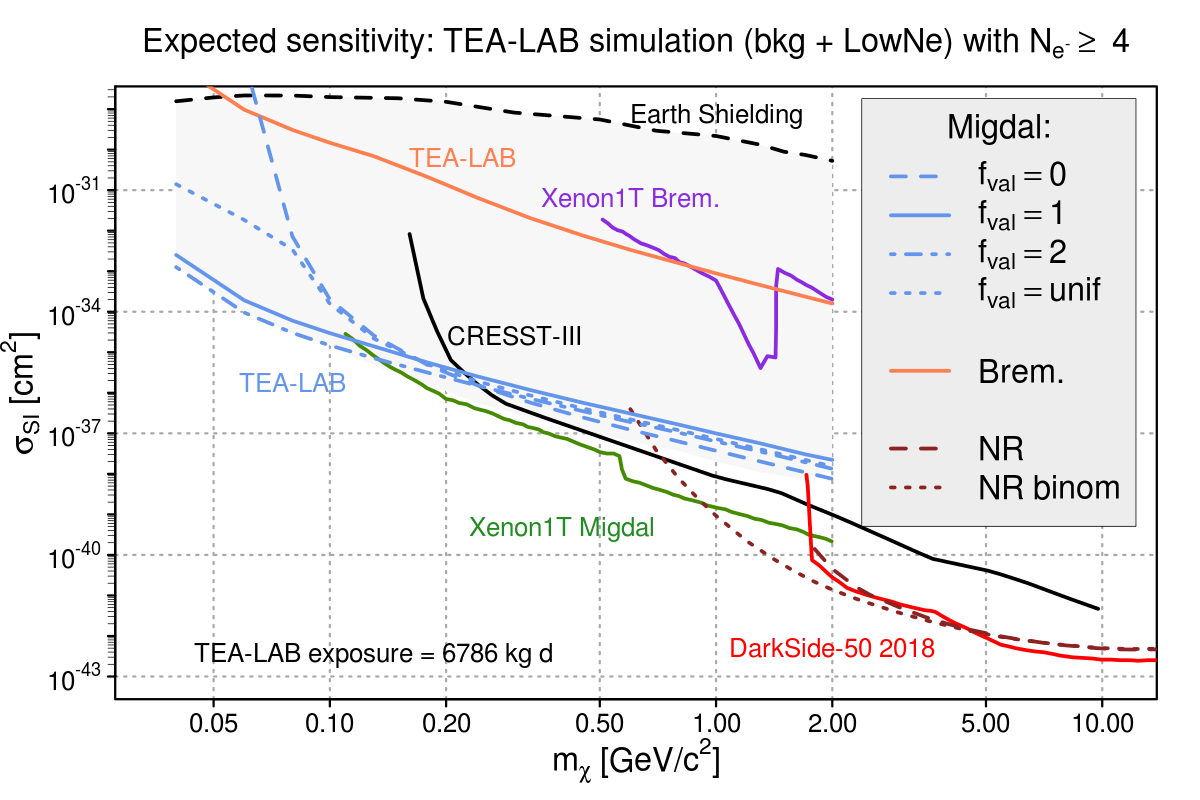}
\caption{$90\%\, {\rm C.I.}$ upper bounds on the $\sigma_{SI}$ exploiting the Migdal electron and photon Bremsstrahlung signals for the {\sc tea-lab} simulated experiment loosely inspired by \DS. These results are obtained using and exposure $E= 6786\,{\rm kg\, d}$, and simulating a pseudo-dataset with the background-only template, including the $LowNe$ excess (see Sec.~\ref{sec:tealab},~\ref{sec:upper_bounds_exp_sensitivity}). The  {\sc tea-lab} bounds are computed for a threshold $N_{e^-} = 4$. Different Migdal electron signal uncertainties are considered (see Sec.~\ref{sec:syst}): $f_{val}=0$ corresponding to no valence shell contribution (blue dashed line), $f_{val}=1$ (blue line), $f_{val}=2$ (dashed-dotted line), and $f_{val}$ treated as a nuisance parameter with a uniform prior p.d.f. in $[0,2]$. The  {\sc tea-lab} bound for the photon Bremsstrahlung signal is reported (orange line).
The estimate of the Earth shielding effect for {\sc tea-lab} is also reported (black dashed curve). The gray shaded area, has to be intended as the region where {\sc tea-lab} has the sensitivity to exclude a DM signal exploiting the Migdal electron. For $m_{\chi} < 0.04{\rm\:GeV}$ we completely loose sensitivity because the signal template is always under threshold.
The upper limits of the Xenon1T~\cite{Aprile:2019jmx}, CRESST~\cite{Abdelhameed:2019hmk}, and \DS~\cite{Agnes:2018ves} are  reported. As a cross check of the {\sc tea-lab} simulation we report also our calculation for the \DS\ bounds on the NR signal with   binomial fluctuation  (brown dotted line) and without (brown dashed line).}
\label{fig:data-bin-4}
\end{center}
\end{figure}

\begin{figure}[!t]
\begin{center}
 \includegraphics[width=.95\textwidth]{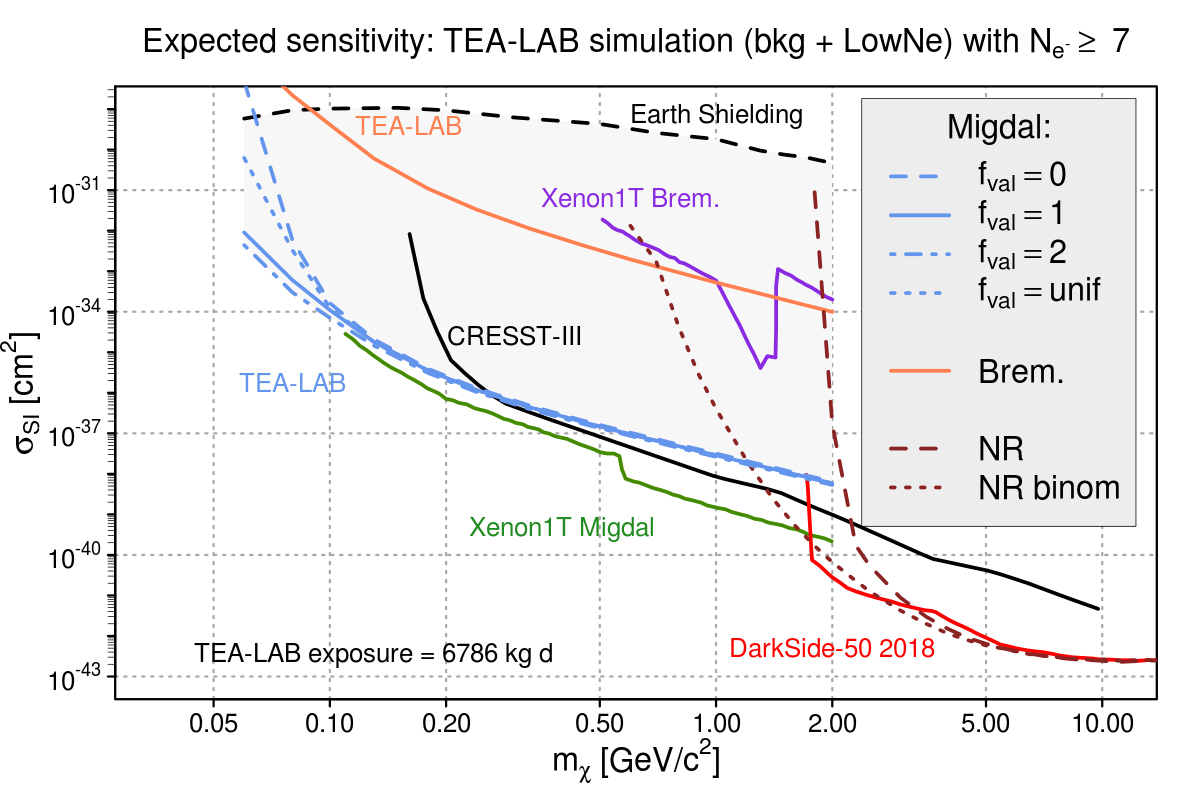}
\caption{$90\%\, {\rm C.I.}$ upper bounds on the $\sigma_{SI}$ exploiting the Migdal electron and photon Bremsstrahlung signals for the {\sc tea-lab} simulated experiment loosely inspired by \DS. These results are obtained using and exposure $E= 6786\,{\rm kg\, d}$, and simulating a pseudo-dataset with the background-only template, including the $LowNe$ excess (see Sec.~\ref{sec:tealab},~\ref{sec:upper_bounds_exp_sensitivity}). The  {\sc tea-lab} bounds are computed for a threshold $N_{e^-} = 7$. Different Migdal electron signal uncertainties are considered (see Sec.~\ref{sec:syst}): $f_{val}=0$ corresponding to no valence shell contribution (blue dashed line), $f_{val}=1$ (blue line), $f_{val}=2$ (dashed-dotted line), and $f_{val}$ treated as a nuisance parameter with a uniform prior p.d.f. in $[0,2]$. The  {\sc tea-lab} bound for the photon Bremsstrahlung signal is reported (orange line).
The estimate of the Earth shielding effect for {\sc tea-lab} is also reported (black dashed curve). The gray shaded area, has to be intended as the region where {\sc tea-lab} has the sensitivity to exclude a DM signal exploiting the Migdal electron. For $m_{\chi} < 0.06{\rm\:GeV}$ we completely loose sensitivity because the signal template is always under threshold.
The upper limits of the Xenon1T~\cite{Aprile:2019jmx}, CRESST~\cite{Abdelhameed:2019hmk}, and \DS~\cite{Agnes:2018ves} are  reported. As a cross check of the {\sc tea-lab} simulation we report also our calculation for the \DS\ bounds on the NR signal with   binomial fluctuation  (brown dotted line) and without (brown dashed line).}
\label{fig:data-bin-7}
\end{center}
\end{figure}

In order to validate our analysis, we compute the bounds for the nuclear recoil for a DM mass in the range $1.8$-$15$ GeV/$c^2$ and compare with the published results of the DarkSide collaboration \cite{Agnes:2018ves}.
We compute the bounds with our bayesian approach reporting the $(90\%\, {\rm C.I.})$ lower bound for both thresholds at $N_{e^-}=4$ and $N_{e^-}=7$ electrons.
The result is shown in Fig.~\ref{fig:data-bin-4} and \ref{fig:data-bin-7} (brown dashed line), together with the \DS\ constraint (solid red line). Here we also validate our simplistic binomial model (brown dotted line) for the experimental response fluctuation.
Our result is in good agreement with the experimental bounds,
considering also the fact that the latter are calculated using the frequentist approach known as CL$_s$ \cite{Cowan:2010js}.

Fig.~\ref{fig:data-bin-4} and \ref{fig:data-bin-7} show the impact of the Migdal effect and the photon Bremsstrahlung for  {\sc tea-lab} in extending the sensitivity region of LAr experiments from $m_{\chi}\sim2{\rm\:GeV}/c^2$ to $m_{\chi}\sim0.05$-$0.08{\rm\:GeV}/c^2$.
For $m_{\chi} \sim 1{\rm\:GeV}/c^2$ the sensitivity based on the Migdal effect is $\sigma_{SI} \sim 10^{-37}{\rm\:cm^2}$; for $m_{\chi} \lesssim 0.11{\rm\:GeV}/c^2$ the sensitivity is comparable with the Xenon1T exclusion limits~\cite{Aprile:2019jmx} and extends up to  $m_{\chi} \sim 0.05$-$0.08{\rm\:GeV}/c^2$.
For the photon Bremsstrahlung the sensitivity spans in the range $m_{\chi} \sim 0.12$-$2{\rm\:GeV}/c^2$, exploring a region inaccessible to Xenon1T.

In addition, we estimate the effect of the Earth attenuation, as discussed in Section~\ref{sec:Earth_attenuation}. We assume for the {\sc tea-lab} estimate a shielding due to the Earth crust compatible with the LNGS underground cavern. 
Our upper limit estimate corresponds to the largest cross section that can be probed by {\sc tea-lab} using only kinematics assumptions. This implies that after a thorough accounting of background contributions, experimental effects, and signal time modulation, the corresponding upper bound might result below the one quoted here. In particular, we computed the bound for the two thresholds under consideration, shown as a black dashed curve in Figures \ref{fig:data-bin-4}-\ref{fig:pseudodata-bin-7}, \ref{fig:data-bkg} and \ref{fig:pseudodata-bin4-20ton}. In these figures the gray shaded area, defined from below by the Migdal $90\%\, {\rm C.I.}$ upper bound, and from above by the Earth attenuation lower bound, has  to be intended as the region where a generic LAr experiment as simulated in {\sc tea-lab} has the sensitivity to exclude a DM signal exploiting the Migdal signal.\footnote{We point out that, since the Earth attenuation lower bound is an estimate based on kinematics consideration, the meaning of the lower and upper bound is different, and thus no clear probability interpretation can be attributed to the sensitivity region depicted in gray.} 

In the next subsections we discuss in details our results exploring the impact of the various systematic effects.

\subsection{Impact of theoretical uncertainties}
\label{sec:expectedsensitivity}

As a first step of our study, we focused on the determination of the impact of the theoretical uncertainties on the outermost shell contribution to the Migdal cross section. As already explained in Sec.~\ref{sec:syst}, since these are $\mathcal{O} (1)$ uncertainties, we decided to compute the limit in four possible scenarios: assuming $f_{val} = {0,1,2}$, or considering $f_{val}$ as a nuisance parameter with a flat prior in the range $[0,2]$. In order to isolate this effect, we produced a {\sc tea-lab} simulated dataset based on the background-only template, without the $LowNe$ excess. 

\begin{figure}[!t]
\begin{center}
 \includegraphics[width=.95\textwidth]{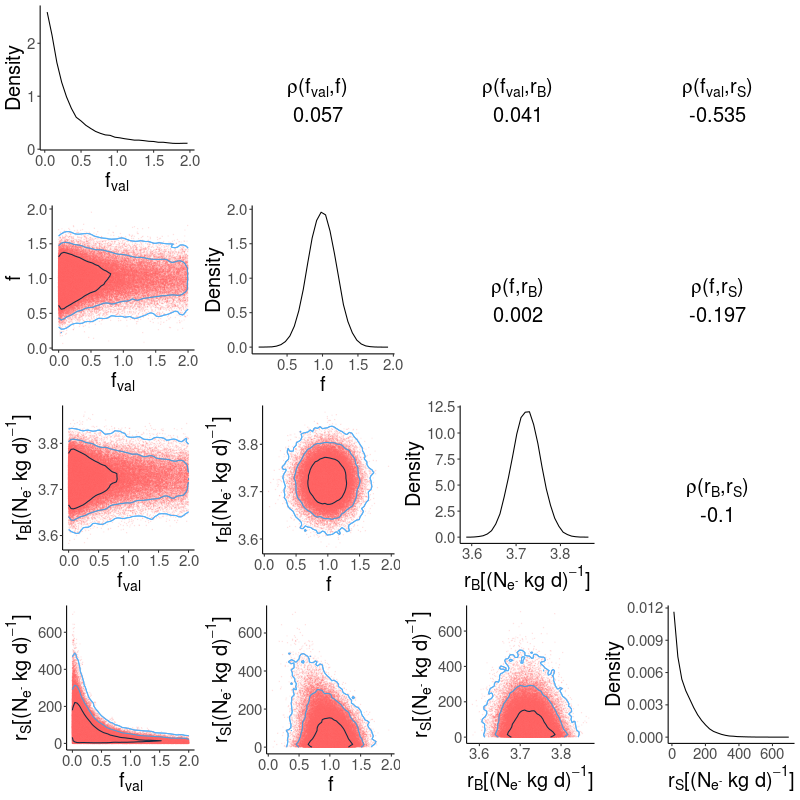}
\caption{Posterior p.d.f. for the relevant parameters of the likelihood using a Migdal-only signal with $m_{\chi} = 130{\rm\:MeV}/c^2$ and a simulated dataset based on the background-only template, without the $LowNe$ excess. The 4 plots on the diagonal of the figure are the uni-dimensional p.d.f. of each single parameter obtained by marginalising on all the others. The 6 bi-dimensional p.d.f. in the bottom-left corner of the figure give the joint p.d.f. of each pair of parameters obtained by marginalising on the others. The plots show also the credible regions at $68\%,\:95\%,\:99.7\%$ probability as solid contour lines. The correlation coefficients are given in the upper-right corner of the figure.}
\label{fig:correlation}
\end{center}
\end{figure}

To make sure that the introduction of additional nuisance parameters does not create instabilities in the fit, we studied the global posterior p.d.f. which is represented in Fig.~\ref{fig:correlation}. This figure, for a mass $m_{\chi} = 130{\rm\:MeV}/c^2$, shows the marginal p.d.f. for each variable as well as a joint p.d.f. for each pair of variables. The posterior p.d.f. looks as expected with the background normalization $r_B$ centered on it's expected value with a normal p.d.f., the parameter of interest $r_S$ has an exponential p.d.f. compatible with a no-signal observation, nuisance parameter $f$ doesn't show any pulls from the input values, while $f_{val}$ is as expected strongly (anti)correlated with $r_S$ with a correlation coefficient $\rho (f_{val}, r_S) = -0.535$.

\begin{figure}[!t]
\begin{center}
 \includegraphics[width=.95\textwidth]{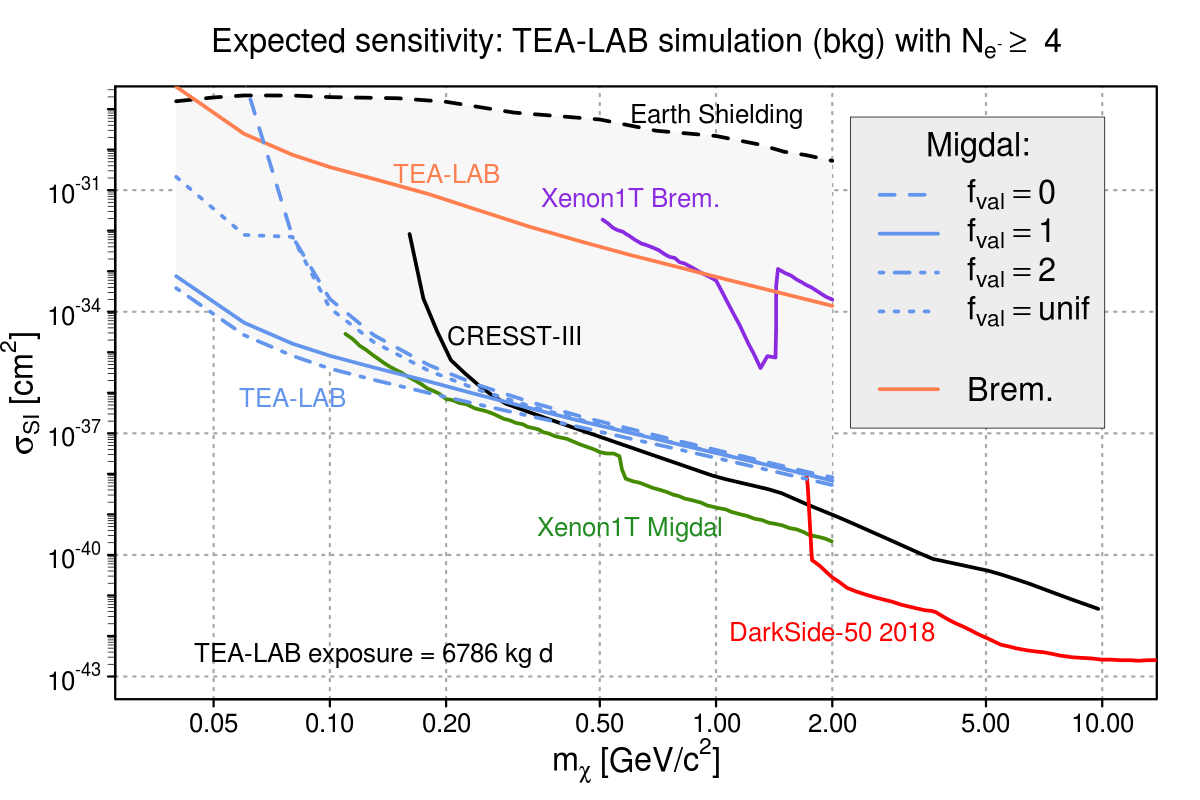}
\caption{$90\%\, {\rm C.I.}$ upper bounds on the $\sigma_{SI}$ exploiting the Migdal electron and photon Bremsstrahlung signals for the {\sc tea-lab} simulated experiment loosely inspired by \DS. These results are obtained using and exposure $E= 6786\,{\rm kg\, d}$, and simulating a pseudo-dataset with the background-only template, not including the $LowNe$ excess (see Sec.~\ref{sec:tealab},~\ref{sec:upper_bounds_exp_sensitivity}).  
The  {\sc tea-lab} bounds are computed for a threshold $N_{e^-} = 4$. Different Migdal electron signal uncertainties are considered (see Sec.~\ref{sec:syst}): $f_{val}=0$ corresponding to no valence shell contribution (blue dashed line), $f_{val}=1$ (blue line), $f_{val}=2$ (dashed-dotted line), and $f_{val}$ treated as a nuisance parameter with a uniform prior p.d.f. in $[0,2]$ (dotted blue line). The  {\sc tea-lab} bound for the photon Bremsstrahlung signal is reported (orange line).
The estimate of the Earth shielding effect for {\sc tea-lab} is also reported (black dashed curve). The gray shaded area, has to be intended as the region where {\sc tea-lab} has the sensitivity to exclude a DM signal exploiting the Migdal electron. For $m_{\chi} < 0.04{\rm\:GeV}$ we completely loose sensitivity because the signal template is always under threshold.
The upper limits of the Xenon1T~\cite{Aprile:2019jmx}, CRESST~\cite{Abdelhameed:2019hmk}, and \DS~\cite{Agnes:2018ves} are  reported.}
\label{fig:pseudodata-bin-4}
\end{center}
\end{figure}

\begin{figure}[!ht]
\begin{center}
 \includegraphics[width=.95\textwidth]{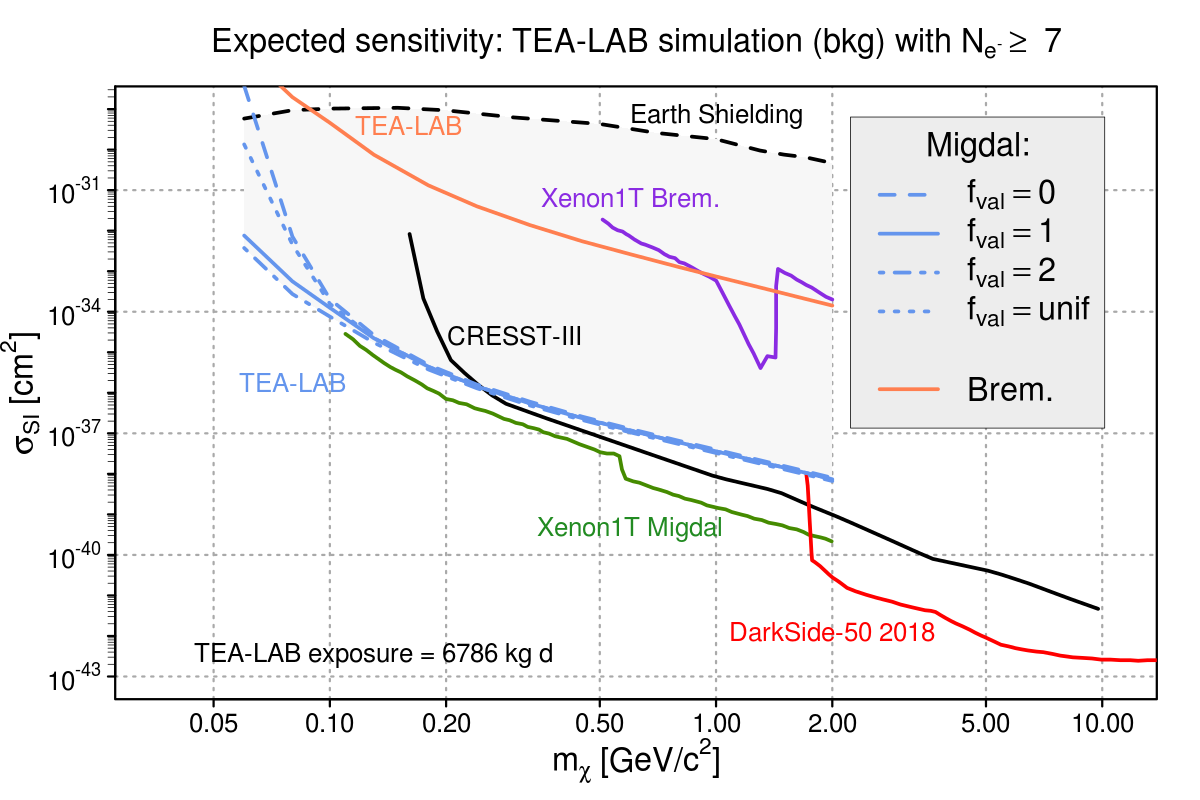}
\caption{$90\%\, {\rm C.I.}$ upper bounds on the $\sigma_{SI}$ exploiting the Migdal electron and photon Bremsstrahlung signals for the {\sc tea-lab} simulated experiment loosely inspired by \DS. These results are obtained using and exposure $E= 6786\,{\rm kg\, d}$, and simulating a pseudo-dataset with the background-only template, not including the $LowNe$ excess (see Sec.~\ref{sec:tealab},~\ref{sec:upper_bounds_exp_sensitivity}). The  {\sc tea-lab} bounds are computed for a threshold $N_{e^-} = 7$. 
Different Migdal electron signal uncertainties are considered (see Sec.~\ref{sec:syst}): $f_{val}=0$ corresponding to no valence shell contribution (blue dashed line), $f_{val}=1$ (blue line), $f_{val}=2$ (dashed-dotted line), and $f_{val}$ treated as a nuisance parameter with a uniform prior p.d.f. in $[0,2]$ (dotted blue line). The  {\sc tea-lab} bound for the photon Bremsstrahlung signal is reported (orange line).
The estimate of the Earth shielding effect for {\sc tea-lab} is also reported (black dashed curve). The gray shaded area, has to be intended as the region where {\sc tea-lab} has the sensitivity to exclude a DM signal exploiting the Migdal electron. For $m_{\chi} < 0.06{\rm\:GeV}$ we completely loose sensitivity because the signal template is always under threshold.
The upper limits of the Xenon1T~\cite{Aprile:2019jmx}, CRESST~\cite{Abdelhameed:2019hmk}, and \DS~\cite{Agnes:2018ves} are  reported.}
\label{fig:pseudodata-bin-7}
\end{center}
\end{figure}

\begin{figure}[t]
\begin{center}
 \includegraphics[width=.95\textwidth]{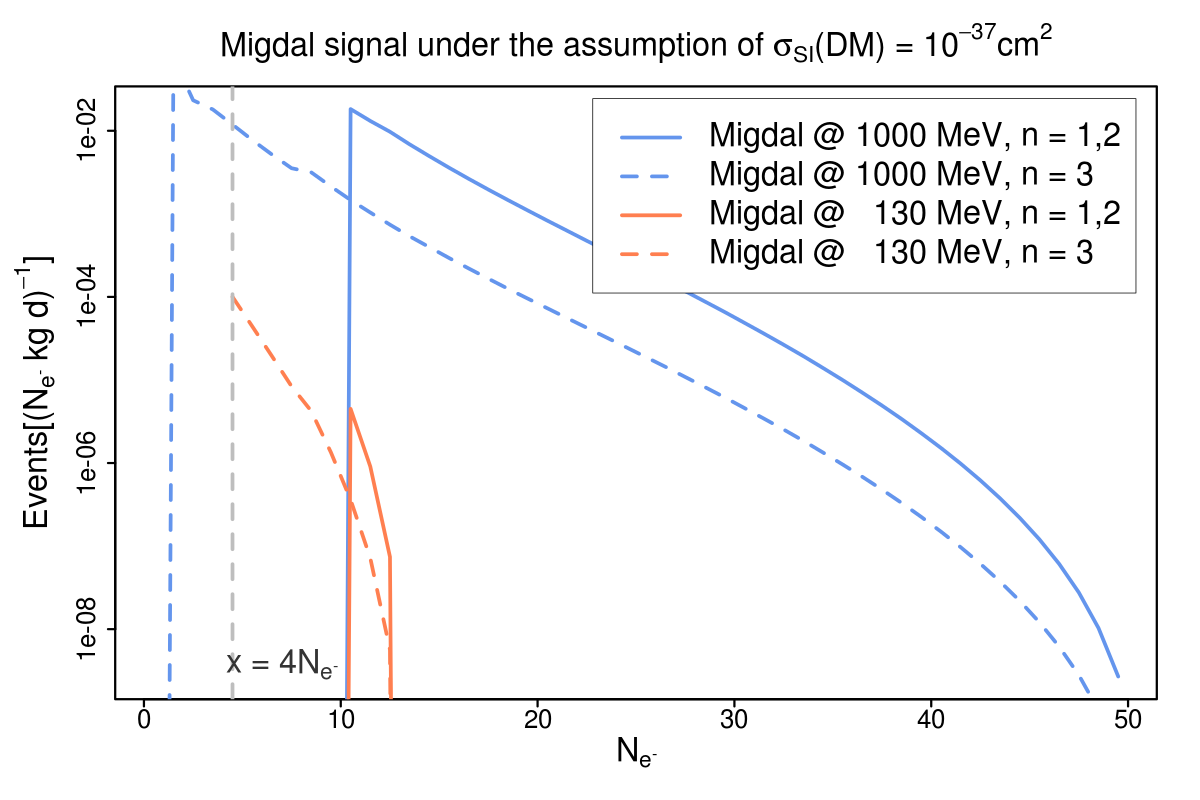}
\caption{Migdal effect signal for $m_{\chi} = 1{\rm\:GeV/c^2}$ (blue lines) and $m_{\chi} = 130{\rm\:MeV/c^2}$ (orange lines) and $\sigma = 10^{-37} {\rm\:cm^2}$,  with $n = 1,2$ (continuous lines) and $n = 3$ (dashed lines).}
\label{fig:signal_templates}
\end{center}
\end{figure}

The sensitivity results are reported in Fig.~\ref{fig:pseudodata-bin-4} and \ref{fig:pseudodata-bin-7} where we considered spectra starting from $N_{e^-} = 4$ and $N_{e^-} = 7$, respectively. 
With respect to sensitivity with $N_{e^-} = 4$ threshold, for $m_{\chi}\lesssim200{\rm\:MeV}/c^2$ the effect of the valence shell is most clear: if $f_{val} \ll 1$ the sensitivity is lost due to the fact that for small masses the greatest contribution to the Migdal signal comes from the outer shell, as depicted in Fig.~\ref{fig:signal_templates}, where we show the Migdal signal for a mass of $130{\rm\:MeV}/c^2$ with respect to the Migdal signal for a mass of $1{\rm\:GeV}/c^2$. As $f_{val}$ increases, the contribution of the third shell becomes more important, and therefore the bound becomes much stronger, even of an order of magnitude. For $m_{\chi}\gtrsim200{\rm\:MeV}/c^2$ this is no more true because the contribution of the inner shells are now of the same intensity of the contribution of the outer shell. Then, for $m_{\chi}\gtrsim200{\rm\:MeV}/c^2$, we can assert that our limit is solid against systematic uncertainties on the contribution coming from the valence shell to Migdal signals in LAr. For the same reasons, if we consider spectra starting from $N_{e^-} = 7$ the contribution of the outer shell is almost completely cut off, and the dependence on $f_{val}$ is weakened, leading to a departure of the sensitivity in the various cases from $m_{\chi}\lesssim100{\rm\:MeV}/c^2$.
With respect to the photon Bremsstrahlung signal, the comparison between the two figures shows that lowering the threshold does not produce a significant change in the sensitivity.


The same considerations hold for Fig.~\ref{fig:data-bin-4},~\ref{fig:data-bin-7} in which the overall sensitivity is reduced, due to the presence of the $LowNe$ excess,
by a factor as big as $\sim 3$.

\subsection{Impact of the experimental effects}
\label{sec:DSobservedsensitivity}

\begin{figure}[!ht]
\begin{center}
 \includegraphics[width=0.95\textwidth]{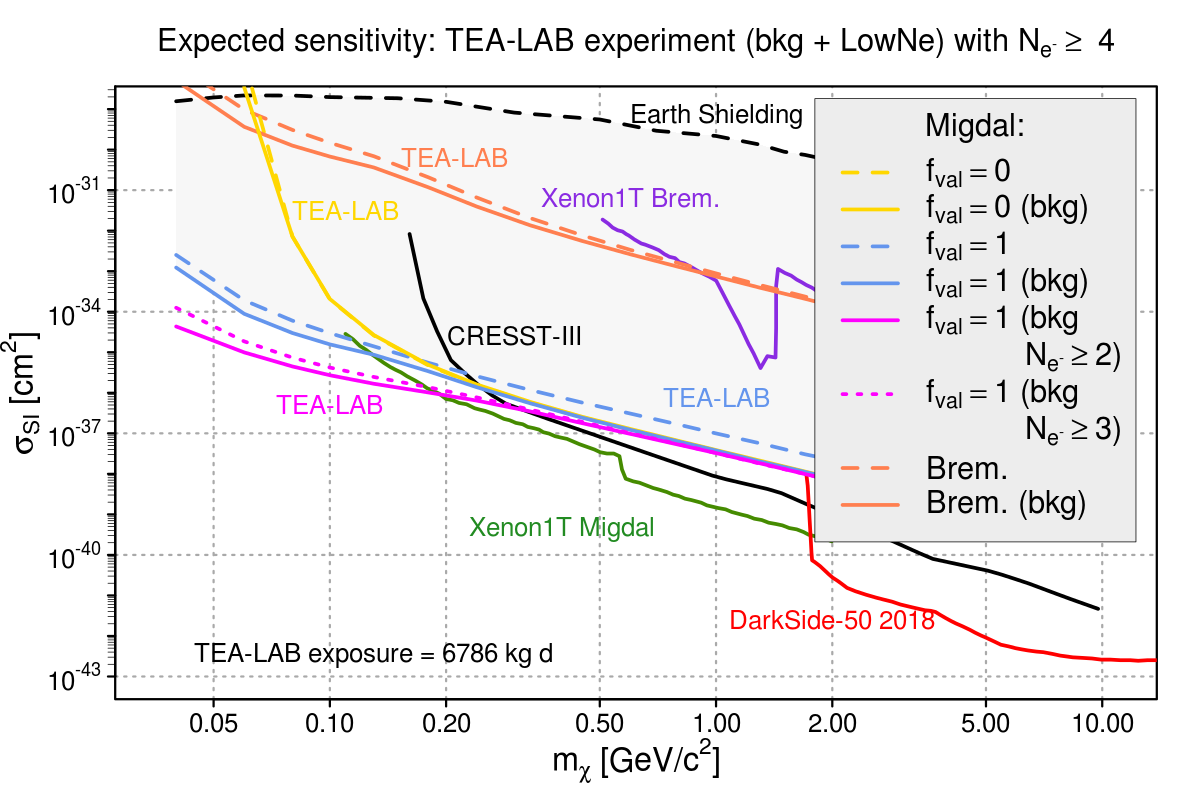}
\caption{$90\%\, {\rm C.I.}$ upper bounds on the $\sigma_{SI}$ exploiting the Migdal electron and photon Bremsstrahlung signals for the {\sc tea-lab} simulated experiment loosely inspired by \DS. These results are obtained using and exposure $E= 6786\,{\rm kg\, d}$, and simulating a pseudo-dataset with the background-only template, including the $LowNe$ excess (see Sec.~\ref{sec:tealab},~\ref{sec:upper_bounds_exp_sensitivity}). The  {\sc tea-lab} bounds are computed for a threshold $N_{e^-} = 4$. 
The solid lines are obtained assigning the $LowNe$ excess to the background as explained in Sec.~\ref{sec:syst}, while the dashed lines give the limit in the default  scenario where the excess is not accounted for as background.
The impact of this effect is given for 2 different Migdal electron signal uncertainties (see Sec.~\ref{sec:syst}): $f_{val}=0$ corresponding to no valence shell contribution (yellow lines) and $f_{val}=1$ (blue lines). For $f_{val} = 1$ we also show the result of considering $N_{e^-} \geq 2$ (magenta solid line) and $N_{e^-} \geq 3$ (magenta dotted line), with the excess accounted for as background.
The estimate of the Earth shielding effect for {\sc tea-lab} is also reported (black dashed curve). The gray shaded area, has to be intended as the region where {\sc tea-lab} has the sensitivity to exclude a DM signal exploiting the Migdal electron. For $m_{\chi} < 0.04{\rm\:GeV}$ we completely loose sensitivity because the signal template is always under threshold.
The upper limits of the Xenon1T~\cite{Aprile:2019jmx}, CRESST~\cite{Abdelhameed:2019hmk}, and \DS~\cite{Agnes:2018ves} are reported.}
\label{fig:data-bkg}
\end{center}
\end{figure}

As we have seen in the previous paragraph a threshold $N_{e^-}=4$ allows to get an expected sensitivity to masses down of $0.06$~GeV/$c^2$ which is much stronger that the one obtained with a higher threshold. To study the impact of $LowNe$ excess in more specific way, as detailed in Sec.~\ref{sec:syst}, we used two strategies. The most conservative is to let the fit account for the excess with a DM signal in the limit calculation procedure. This results in a lower bound on DM and corresponds to the blue lines in Fig.~\ref{fig:data-bin-4}. We produce results also in the case where the excess would be understood and modeled as a background. To emulate this circumstance we fit the excess as described in Sec.~\ref{sec:syst} and assign it to the background component. The new bounds are plotted in Fig.~\ref{fig:data-bkg} as a solid blue line (for $f_{val}=1$, i.e. including the contribution of the valence shell) and yellow line (for $f_{val}=0$). For comparison the dashed lines give the limit in the default scenario where the excess is not accounted for as a background. We notice that, if the contribution of the valence electrons is neglected, there is no difference between the two approaches. This is because the inner shells do not give any signal in the region where the excess is present (see Fig.~\ref{fig:signal_templates}). On the contrary, since the valence shell contribute significantly to the region where the excess appears, accounting for it with an additional background contribution leads to a stronger bound, roughly independent of the mass.
In Fig.~\ref{fig:data-bkg} we also show how the sensitivity bounds would improve extending our analysis to $N_{e^-} \geq 3$ and $N_{e^-} \geq 2$, always assigning the $LowNe$ excess to the background component.
Finally, as for Sec.~\ref{sec:expectedsensitivity}, for a threshold $N_{e^-} = 7$ the dependence on $f_{val}$ is weakened.

\begin{figure}[th]
\begin{center}
 \includegraphics[width=.95\textwidth]{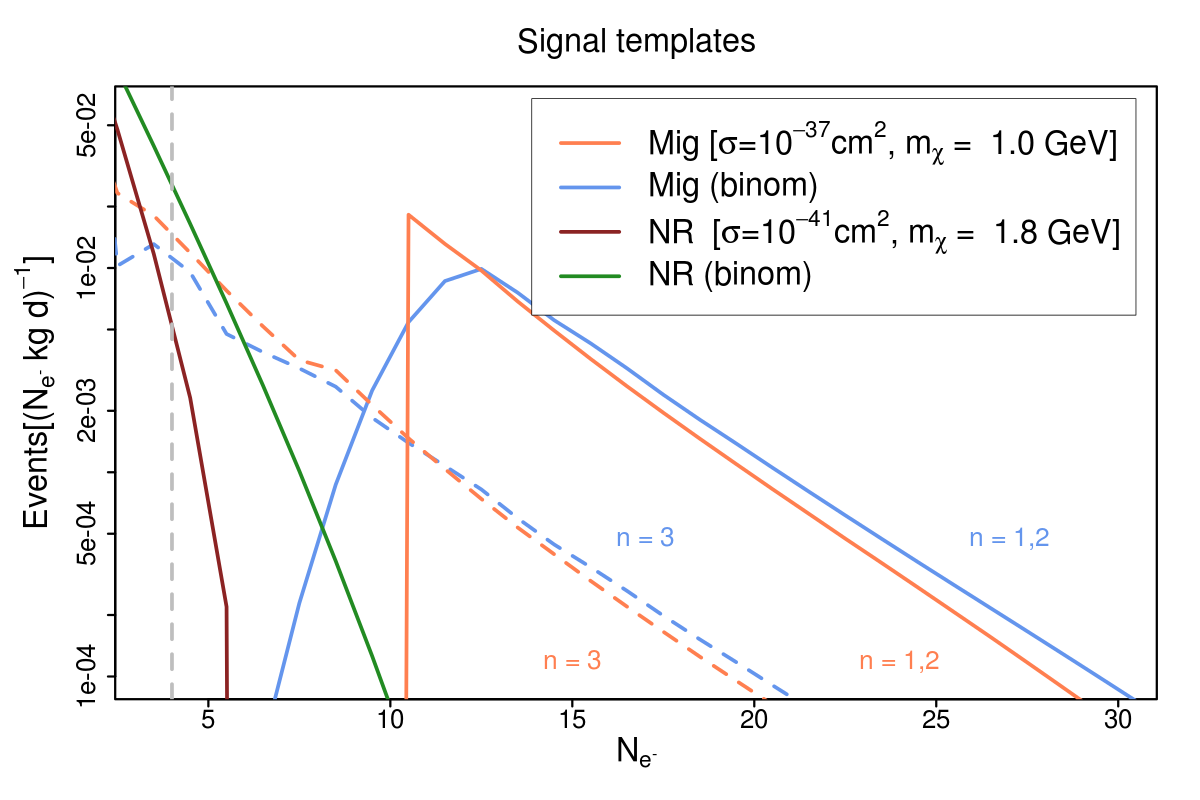}
\caption{Migdal electron signal for $m_{\chi} = 1{\rm\:GeV/c^2}$ (blue lines) and $m_{\chi} = 1{\rm\:GeV/c^2}$ folded with the binomial model for the detector response (orange lines), both evaluated for $\sigma = 10^{-37} {\rm\:cm^2}$, considering only $n=1,2$. The dashed lines represent the same signals including also  $n = 3$. NR for $m_{\chi} = 1.8{\rm\:GeV/c^2}$ and $\sigma = 10^{-41} {\rm\:cm^2}$
with the binomial model for the detector response (green line), without (brown line).}
\label{fig:signal_templates_bin}
\end{center}
\end{figure}

Fig.~\ref{fig:signal_templates_bin} shows also the impact on the signal spectra due to the binomial fluctuation of the detector response. For a NR signal at $m_{\chi} = 1.8{\rm\:GeV/c^2}$, it is clearly visible how these fluctuations let some of the events spill above the $N_{e^-}=4$ threshold. This effect is not present for the Migdal electron spectra since they extend well above the analysis threshold independently of the fluctuations model. We checked that including the binomial model does not change the results for the Migdal electron and photon Bremsstrahlung.

In conclusion, we point out that to exploit at the best the contribution of the Migdal effect and the photon Bremsstrahlung, it is crucial to have a precise description of the backgrounds in the low $N_{e^-}$ region.

\subsection{Projected sensitivity for future experiments}
\begin{figure}[!ht]
\begin{center}
 \includegraphics[width=.95\textwidth]{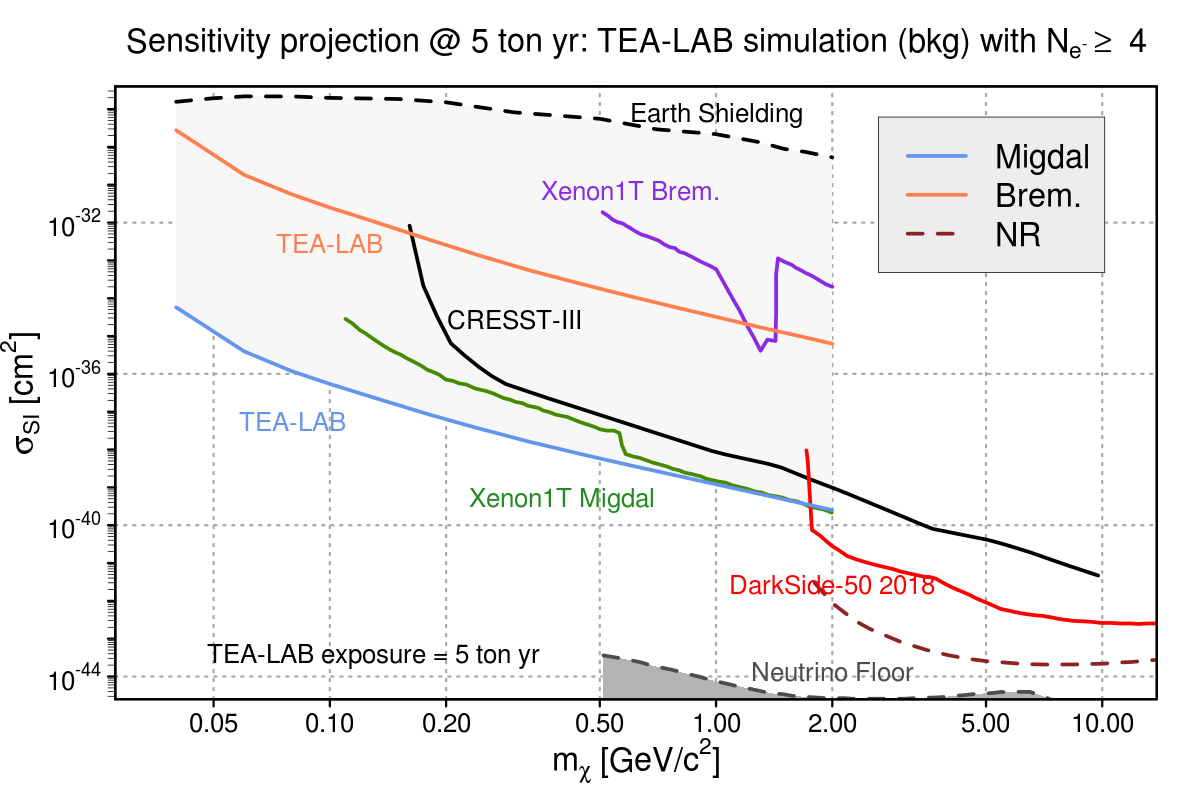}
\caption{$90\%\, {\rm C.I.}$ upper bounds on the $\sigma_{SI}$ exploiting the Migdal electron and photon Bremsstrahlung signals for the {\sc tea-lab} simulated experiment loosely inspired by \DS. These results are obtained using and exposure $E= 5\,{\rm ton\, yr}$, and simulating a pseudo-dataset with the background-only template, not including the $LowNe$ excess (see Sec.~\ref{sec:tealab},~\ref{sec:upper_bounds_exp_sensitivity}). The  {\sc tea-lab} bounds are computed for a threshold $N_{e^-} = 4$. 
The estimate of the Earth shielding effect for {\sc tea-lab} is also reported (black dashed curve). The gray shaded area, has to be intended as the region where {\sc tea-lab} has the sensitivity to exclude a DM signal exploiting the Migdal electron. For $m_{\chi} < 0.04{\rm\:GeV}$ we completely loose sensitivity because the signal template is always under threshold.
The upper limits of the Xenon1T~\cite{Aprile:2019jmx}, CRESST~\cite{Abdelhameed:2019hmk}, and \DS~\cite{Agnes:2018ves} are  reported. We report also our calculation for the \DS\ bounds on NR signal extrapolated at $E= 5\,{\rm ton\, yr}$ (brown dashed line). The neutrino floor for LAr experiments \cite{Agnes:2018ves} is reported as a dark-gray shaded area.}
\label{fig:pseudodata-bin4-20ton}
\end{center}
\end{figure}

To illustrate how the sensitivity to DM particles enhanced by the Migdal electron and the photon Bremsstrahlung scales with the experimental exposure, we show in Fig.~\ref{fig:pseudodata-bin4-20ton} the {\sc tea-lab} 90\%~C.I. bounds for an exposure $E=5\, {\rm ton\, yr}$. This exposure corresponds roughly to an increase of a factor $RE=270$ with respect to the results presented in Fig.~\ref{fig:pseudodata-bin-4} for the exposure $E=6786\, {\rm kg~d}$. As expected, the improving factor in the sensitivity is about $1/\sqrt{RE}\simeq 16$, which corresponds to the square root of the increased exposure. For this projection we also scaled the uncertainties by a factor of  $1/\sqrt{RE}$ in order not to be dominated by systematic effects. We point out that the best achievable improvement with the exposure is attainable only if the bounds are limited by the sample size. When the statistical uncertainty becomes comparable with the systematic uncertainty, the limit stops improving with an augmented data sample. This effect is particularly relevant for the background subtraction. In fact, already with an exposure factor $RE\gtrsim10$, the limits become dominated by systematic effects due to the large contribution of the backgrounds and their limited knowledge. This effect can be canceled up to a certain $RE$ by having a more precise knowledge of the background contamination; however, already at $RE=270$, the background uncertainty has to be smaller than 1\textperthousand.
This clearly imposes the necessity to define strategies to reduce the background contamination to push the limit down with large exposures. Fig.~\ref{fig:pseudodata-bin4-20ton} shows also the neutrino floor for a liquid argon experiment as given in ref.~\cite{Agnes:2018ves}. For $m_{\chi}\sim 6{\rm\:GeV}/c^2$ the projected sensitivity for the nuclear recoil spin independent signal starts to approach the neutrino floor. For the Migdal effect this is not the case as we are few order of magnitude above. However, we point out that it will become important to have a reliable estimate of the neutrino floor in liquid argon experiments for masses below $m_{\chi}\lesssim 0.5{\rm\:GeV}/c^2$.

Bearing these considerations in mind, it appears clear that any reasonable extrapolation of our result to higher exposure as for example DS20k~\cite{Aalseth:2017fik}, which is foreseen reaching values as big as $E=3000\, {\rm ton\, yr}$ with a completely redesigned detector, can only be done by the experimental collaboration after a thorough assessment of the relevant systematic effects.

\clearpage
\section{Conclusions}
\label{sec:conclusions}

We considered both the Migdal effect and photon Bremsstrahlung from the nucleus hit by a DM candidate in experiments based on LAr detectors. Previous DM searches exploiting LAr detectors include \DS\ \cite{Agnes:2018fwg} and DEAP-3600 \cite{Ajaj:2019imk}, and have mainly focused on masses greater than $10$~GeV/$c^2$ by considering only the nuclear recoil signal.  We decided to develop the {\sc tea-lab} simulated experiment taking inspiration from the \DS\ experiment, for which there is a published low-mass analysis \cite{Agnes:2018ves} sensitive down to masses of $1.8$~GeV/$c^2$. Since at low masses relative large DM cross sections are probed, we estimated the effect of the Earth attenuation, which results in a lower limit on $\sigma_{SI}$ above which the experiment becomes blind.

In order to study the sensitivity to low masses exploiting the Migdal effect and photon Bremsstrahlung, we built a simplified likelihood describing the experimental response of {\sc tea-lab}. In particular we payed attention to background components, to the description of the low $N_{e^-}$ excess, and to the detector response including the fluctuation in the number of ionization electrons. We also considered the systematic uncertainties on the theoretical computation of the Migdal spectra, among which the contribution of the valence shell of argon plays a critical role, being estimated with an uncertainty of $\mathcal{O}(1)$. We therefore used this likelihood in the context of a Bayesian approach to compute  the expected sensitivity, the exclusion limits at the $90\%{\rm\:C.I.}$, and the projections at higher exposure. We show that the limits are mildly dependent on reasonable prior choices describing the present state of knowledge on the sought signals. 

As a preliminary cross check, we  computed the limit for the nuclear recoil signal using the \DS\ measured spectrum obtaining a result compatible with the expected one.

We then studied the {\sc tea-lab} expected sensitivity at the \DS\ exposure and show the impact of the systematic effects on the final result. The sensitivity spans masses in the range~$0.1$-$2{\rm\:GeV}/c^2$. For $m_{\chi}\gtrsim0.2{\rm\:GeV}/c^2$, the results are stable and insensitive to significant theoretical uncertainties on the argon valence shell contribution to the Migdal effect. For $m_{\chi}\lesssim0.2{\rm\:GeV}/c^2$ instead these effects can substantially modify the limit, even by two orders of magnitude for the lowest masses (see Fig.~\ref{fig:pseudodata-bin-4}).

We studied the interplay between the poorly understood $LowNe$ excess and the Migdal effect showing that the bounds change only if the contribution of the valence shell is considered. In this case, a better understanding of the excess results in a stronger limit. 
For $m_{\chi}\sim0.11{\rm\:GeV}/c^2$ the limit is comparable with that of the Xenon1T experiment obtained with a much larger exposure, and it reaches  $m_{\chi}\sim0.06{\rm\:GeV}/c^2$ extending the sensitivity of noble liquids to lower masses.

Finally we projected the sensitivity analysis up to an exposure $E=5{\rm\:ton\:yr}$, finding that the current knowledge of the background systematic uncertainties prevents the limit to scale with the statistics for exposure larger than $\sim0.2{\rm\:ton\:yr}$. This shows how any reasonable extrapolation of our results can only be done by the experimental collaboration after a thorough assessment of the relevant systematic effects of the foreseen experiment.

We emphasize that the large uncertainty on the contribution of the valence shell for the computation of the Migdal effect already plays a crucial role in setting a bound from argon detectors for DM masses below $0.2\,\rm{GeV}/c^2$. As a consequence, a dedicated theoretical effort to compute the transition probabilities with higher precision is needed. 

We conclude by stressing that the contribution of a Bremsstrahlung photon and a Migdal electron is sizable also for LAr experiments, and it is a powerful tool to explore mass regions below the GeV$/c^2$ scale, inaccessible by modeling the signal only with the traditional nuclear recoil interaction. We look forward to an update of low masses analyses in LAr experiments that includes such effects.

\section*{Acknowledgements}
It is a pleasure to acknowledge Marco Nardecchia for the stimulating discussions on the topic. We furthermore thank M. Ibe, B. Kavanagh and N. Spaldin for useful correspondence. GGdC has been supported by the National Science Centre, Poland, under research grant no. 2017/26/D/ST2/00225.

\bibliographystyle{JHEP}
\bibliography{migdalLar} 

\end{document}